\documentclass[preprint]{aastex}

\usepackage{natbib}
\usepackage{apjfonts}
\usepackage{aas_macros}
\usepackage{lscape}
\usepackage{url}
\usepackage{rotating}

\shorttitle{NGC 404}
\shortauthors{Boehle et al.}

\begin{document}

\title{Widespread Shocks in the Nucleus of NGC 404\\Revealed by Near-infrared Integral Field Spectroscopy}

\author{A. Boehle\altaffilmark{1,2}, J.~E.~Larkin\altaffilmark{2}, L.~Armus\altaffilmark{3}, S.~A.~Wright\altaffilmark{4,5}}
\altaffiltext{1}{Institute for Particle Physics and Astrophysics, ETH Zurich, CH-8093 Zurich, Switzerland}
\altaffiltext{2}{Department of Physics \& Astronomy, University of California,
    Los Angeles, CA, 90095, USA}
\altaffiltext{3}{IPAC, California Institute of Technology, Pasadena, CA, 91125, USA}
\altaffiltext{4}{Department of Physics, University of California, San Diego, CA, 92039, USA}    
\altaffiltext{5}{Center for Astrophysics and Space Sciences, University of California, San Diego, CA, 92039,
USA}

\begin{abstract}
We present high spatial resolution, integral field spectrograph (IFS) observations of the nearby LINER (low ionization nuclear emission line region) galaxy NGC 404 at 1.25 $\mu$m (J band) and 2.2 $\mu$m (K band) near-infrared wavelengths. 
Although NGC 404 is thought to host an intermediate mass black hole at its center, it has been unclear whether accretion onto the black hole or another mechanism such as shock excitation drives its LINER emission at optical/near-infrared wavelengths.  We use the OSIRIS IFS at Keck Observatory behind laser guide star adaptive optics to map the strength and kinematics of [FeII], H$_2$, and hydrogen recombination lines at spatial resolutions of 1 pc across the central 30 pc of the galaxy.  
The H$_2$ gas is in a central rotating disk and ratios of multiple H$_2$ lines indicate that the molecular gas is thermally excited, with some contribution from UV fluorescence.  The [FeII] emission is more extended and diffuse than the molecular gas and has a different kinematic structure that reaches higher velocities/dispersions.  We also map the strength of the CO stellar absorption feature and constrain the dominant age of the nuclear stellar population to $\sim$1 Gyr.  
Finally, we find regions across the nucleus of NGC 404 with [FeII]/Pa$\beta$ line ratios up to 6.5, $\sim$2.5 times higher than the ratio measured from spatially-integrated spectra.  
From these high line ratios, we conclude that shocks are the dominate physical mechanism exciting NGC 404's LINER emission and argue that a possible source of this shock excitation is a supernova remnant.  

\end{abstract}

\keywords{galaxies: individual (NGC 404) $-$ techniques: imaging spectroscopy $-$ galaxies: nuclei $-$ galaxies: active $-$ infrared: galaxies}

\section{Introduction}
\label{sec:intro}
LINER (low ionization nuclear emission line region) galaxies, first defined by \citet{1980A&A....87..152H}, are the least luminous but most common class of active galactic nucleus (AGN), with $\sim$1/3 of all galaxies within 40 Mpc hosting LINER emission at their centers \citep{1997Ho_PaperV}. 
LINERs are identified by their optical emission line properties and show high values of line ratios with low ionization species (e.g., [NII]/H$\alpha$) similar to Seyferts, but low values of line ratios from high ionization species ([OIII]/H$\beta$).  These optical line ratios can be explained by a number of excitation mechanisms, with two dominant models being photoionization from a weakly accreting supermassive black hole (BH) or fast shock excitation \citep{2013Riffel}.  
Although the physical mechanisms exciting optical LINER emission likely vary across the class, 
studying LINERs in detail can reveal under what circumstances different excitation mechanisms dominate and help us determine the importance of BH accretion, feedback, and other physical processes such as star formation in the nuclei of relatively normal galaxies.  The closest LINERs are ideal targets for detailed studies of individual nuclei because diffraction-limited angular resolutions of $\sim$0.1 arcsec can probe physical scales of 5 pc, allowing the excitation mechanisms in a single nucleus to be disentangled at very high spatial resolutions.

NGC 404 hosts one of the closest LINERs \citep{1997Ho_PaperIII} at a distance of only 3.1 Mpc away \citep{2016ApJ...822...70L} and is therefore an ideal source for studying the important physical processes at the centers of these galaxies.  
There is tentative evidence that NGC 404 hosts an accreting central BH from detections at X-ray \citep{2011ApJ...737...77B} and radio \citep[e.g.,][]{2012ApJ...753..103N} wavelengths and variability observed at near-infrared (NIR), optical, and UV wavelengths \citep{2005ApJ...625..699M,2017ApJ...836..237N}. 
A recent work attributed the radio source to a ratio jet launched by a central low-mass, low-Eddington ratio BH \citep{2017ApJ...845...50N}. 
NIR studies of the molecular gas and stellar dynamics in the nucleus of NGC 404 have placed an upper limit on the mass of this putative central black hole of 1.5 x 10$^5$ M$_{\odot}$, in the mass range of an intermediate mass black hole \citep{2010Seth,2017ApJ...836..237N}. 
Despite the evidence for the presence of an accreting central BH, 
it has been unclear whether photoionization from the BH accretion disk or another mechanism such as shock excitation is the dominant physical process in the NGC 404 nucleus driving its LINER emission.

NGC 404, and LINERs in general, also stand out from other emission line galaxies in their NIR spectral properties.  NIR spectra of these nuclei in the J (1.1-1.4 $\mu$m) and K bands (2.0-2.4 $\mu$m) often show features from [FeII], H$_2$, and the Pa$\beta$ and Br$\gamma$ hydrogen recombination lines.  
At the spatial resolutions of $\sim$100 pc probed by previous seeing-limited slit spectroscopy observations, LINERs exhibit high ratios of [FeII]/Pa$\beta$ and H$_2$/Br$\gamma$ compared to Seyfert and starburst nuclei \citep{1998Larkin, 2004A&A...425..457R, 2005MNRAS.364.1041R, 2013Riffel}. 
NGC 404 in particular shows some of the highest values of these line ratios, with the [FeII]/Pa$\beta$ line ratio of 2.7 being similar to shocked regions of supernova remnants \citep{1998Larkin}.  The [FeII]/Pa$\beta$ ratio is sensitive to shock excitation because iron is typically locked in dust grains, and high values of [FeII]/Pa$\beta$ (> 2.0) require shocks that can destroy or ablate the dust and convert a sufficient amount of iron to a gaseous state \citep{1994Blietz, 1998Larkin}.  
With the advent of adaptive optics (AO), the NIR diagnostic spectral features can be probed in LINERs at spatial resolutions 100 times smaller than seeing-limited observations using NIR integral field spectroscopy (IFS).  These data consist of thousands of spectra taken simultaneous and can resolve differences in the [FeII] and H$_2$ emission and kinematics on spatial scales of only 1 pc at the nearby distance of NGC 404, thereby spatially separating out regions in the nucleus of this galaxy that are dominated by shock excitation, star formation, and other physical processes.  
Using IFS data to spatially resolve differences in the [FeII]/Pa$\beta$ line ratio in particular can reveal regions of high [FeII]/Pa$\beta$ that are averaged out in seeing-limited spectra.  
LINERs in general have been included in past NIR IFS studies of nearby galactic nuclei \citep[e.g.,][]{2013MNRAS.428.2389M, 2013ApJ...763L...1M}, but these studies focused on characterizing the molecular gas emission and the stellar absorption features in the K band and did not probe the shock-sensitive [FeII] emission line in the J band. 

We present results on NIR IFS observations of the nucleus of NGC 404 using the OSIRIS instrument at the W. M. Keck Observatory.  These observations are a part of a larger survey of nearby LINER galaxies with the goal of constraining the excitation mechanisms of NIR emission lines at the highest possible spatial scales.  We present data at both the J and K bands that measure the emission morphology, kinematics, and physical conditions of the ionized iron and molecular hydrogen gas at spatial scales of 1 pc.  This work includes the first high spatial resolution study of the shock-sensitive [FeII] line in NGC 404. Based on the spatially-resolved spectral features and continuum emission, we find that the OSIRIS observations of the NGC 404 nucleus are best explained by at least four major physical components including a nuclear stellar population, a central rotating H$_2$ disk, an HII region, and an extended shock front traced by strong [FeII] emission. 
Sec.~\ref{sec:observations} describes the OSIRIS observations and data reduction procedures.  Sec.~\ref{sec:analysis} describes the data analysis and the resulting measurements of the stellar and ionized/molecular gas properties.  Sec.~\ref{sec:discussion} discusses the implications of these results on the stellar population and on the physical mechanisms exciting the NIR emission in NGC 404.

\section{Observations and Data Reduction}
\label{sec:observations}

NGC 404 was observed on 2 September 2014 and 8 October 2015 using the 
OSIRIS instrument \citep{2006NewAR..50..362L} on the Keck I telescope. 
OSIRIS is an integral field spectrograph coupled to the Keck Adaptive Optics System \citep{2006PASP..118..297W, 2006PASP..118..310V}.  This instrument uses an array of small lenses to sample the AO image and produces up to $\sim$3000 spectra simultaneously across a rectangular, contiguous field of view.  
The data were taken with the 50 mas (September 2014) and 35 mas (October 2015) plate scale modes using two different filters: Jn2 ($\lambda$ = 1.228 - 1.289 $\mu$m) and Kbb ($\lambda$ = 1.965 - 2.381 $\mu$m).  The Jn2 filter covers the [FeII] ($\lambda_{0}$ = 1.2567 $\mu$m) and Pa$\beta$ ($\lambda_{0}$ = 1.28216 $\mu$m) emission lines for low redshift galaxies (z $<$ 0.004).  The Kbb filter covers a series of H$_2$ emission lines, including the 1-0 S(1) transition ($\lambda_0$ = 2.1218 $\mu$m), as well as the Br$\gamma$ line ($\lambda_0$ = 2.16612 $\mu$m) and stellar absorption features, particularly the CO bandheads ($\lambda_0$ = 2.29 $\mu$m).  At the nearby distance of NGC 404, the angular pixel scales of 50 mas and 35 mas correspond to 0.8 pc and 0.5 pc respectively.  See Table \ref{tab:obs} for a summary of the OSIRIS observations of NGC 404.

The on-source observations of NGC 404 total 0.75 hours in the 50 mas mode on 2 September 2014 and 1.25 hours in the 35 mas mode on 8 October 2015 in each filter, with individual frames having an integration time of 900 sec.  The airmass of NGC 404 during the observations ranged from 1.0 - 1.2.   
The data were taken using the Keck laser guide star AO system \citep{2006PASP..118..297W, 2006PASP..118..310V}, with the bright, point-like nucleus of the galaxy serving as the tip/tilt reference source \citep[$V_{\textrm{mag}} \approx 15$ within 0.5 arcsec;][]{2008AJ....135..747G}.  The galaxy was dithered within the field of view such that previous or subsequent on-source frames could be used for sky subtraction, with typical dithers being 1 arcsec.  In addition to the on-source frames, one 900-sec pure sky frame offset $\sim$4 arcseconds from the source was taken in each filter and in each plate scale mode.

\begin{deluxetable}{llrrrrrrr}
\tabletypesize{\scriptsize}
\tablewidth{0pt}
\tablecaption{Summary of OSIRIS Observations of NGC 404}
\tablehead{
  \colhead{} & 
  \colhead{} & 
  \colhead{Wavelengths} & 
  \colhead{Plate Scale} & 
  \colhead{\# of } &
  \colhead{Integration} &
  \multicolumn{2}{c}{FOV (arcsec)} \\
  \colhead{Date} &
  \colhead{Filter} &  
  \colhead{($\mu$m)} &   
  \colhead{(mas)} &   
  \colhead{Frames} &
  \colhead{Time (hours)} & 
  \colhead{Single Frame} & 
  \colhead{Mosaic} & 
}
\startdata
2 September 2014  &  Jn2 & 1.228 - 1.289  &  50 &  3 & 0.75 & 2.1 x 3.2  & 2.1 x 3.2 \\ 
2 September 2014  &  Kbb & 1.965 - 2.381  &  50 &  3 & 0.75 & 0.8 x 3.2 & 0.8 x 3.2 \\
8 October 2015  &  Jn2 & 1.228 - 1.289 &  35 &  5 & 1.25 & 1.47 x 2.24  & 1.65 x 3.29 \\
8 October 2015  &  Kbb & 1.965 - 2.381  &  35 &  5 & 1.25 & 0.56 x 2.24  & 0.70 x 3.33
\enddata 
\label{tab:obs}
\end{deluxetable}

To correct for atmospheric absorption features, we also observed two telluric standard stars at an airmass similar to the science observations and with the same plate scale/filter setups: an A star (A0 star HD 13869 in 2014 and A1 star HD 217186 in 2015) and a G2 star (HD 12846 in 2014 and HD 217577 in 2015).  
Both the A and G star spectra were divided by a blackbody spectrum to remove the continuum shape of the stellar photosphere.  The spectrum of the A star was used as the telluric spectrum across the majority of both observation bands, with the exception of the regions around the strong hydrogen absorption features of Br$\gamma$ and Pa$\beta$.  For these wavelength regions (1.270 - 1.289 $\mu$m in Jn2 and in 2.156 - 2.177 $\mu$m Kbb) the G2 spectrum was used, with the stellar absorption features removed by dividing the G2 spectrum by the solar spectrum.  
The solar spectrum used to correct the G2 star was created by ESO using National Solar Observatory/Kitt Peak Fourier Transform Spectrograph data produced by NSF/NOAO\footnote{\url{http://www.eso.org/sci/facilities/paranal/decommissioned/isaac/tools/spectroscopic_standards.html}}.  
Before dividing the OSIRIS G2 spectrum by the solar spectrum, the solar spectrum was shifted to match the radial velocity (RV) of the star and scaled so the width of the hydrogen absorption line matched the G2 spectrum.  Note that the value of the blackbody temperature used correct for the shape of the stellar continuum emission affects the final telluric spectrum at a level of <1\% for a wide range of temperatures ($\pm$2000 K). 
Finally, the resulting telluric spectrum was normalized to a median of 1.0 across all wavelengths, so that dividing by this spectrum corrects for the atmospheric absorption and instrumental throughput as a function of wavelength and does not change the overall level of the galaxy spectrum.

All the data were reduced using Version 3.2 of the OSIRIS Data Reduction Pipeline \citep[DRP;][]{Krabbe2004}.  Each individual science frame was sky subtracted by both the pure sky and its dithered pair frame (if applicable).  The scaled sky DRP routine was also performed on each frame.  This routine scales the intensity of different families of OH lines in intensity to match the sky frame and the science frame and is based on the algorithm developed by R. Davies \citep{2007MNRAS.375.1099D}.  All sky subtracted frames were visually compared and the frame with the lowest OH line residuals was chosen as the final sky subtracted frame.  These sky-subtracted frames were then mosaicked to create the final data cube for each filter.  These final cubes have a wavelength sampling of 0.00015 $\mu$m in the Jn2 band and 0.00025 $\mu$m in the Kbb band, corresponding to a velocity resolution of $\sim$35 km s$^{-1}$ in both bands.  The reduced data cubes were also wavelength calibrated by the DRP using information from arc lamp calibration data and OH sky lines. Finally, the spectrum of each spatial pixel, or spaxel, in the mosaicked cube 
was divided by the telluric spectrum created from the A and G star spectra. 
All subsequent data analysis was performed on these telluric-corrected data cubes.  The units of the final spectra in each wavelength channel as reduced by the DRP are counts, which are proportional to flux density of the galaxy.

\begin{figure}[t]
	\begin{center}
	\includegraphics[width=13cm]{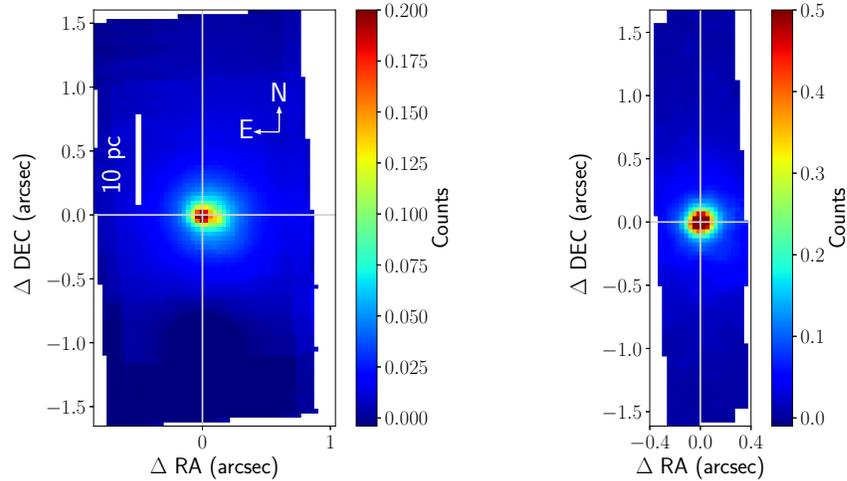}
	\end{center}
	\caption
	{ \label{fig:reducedcubes} The NGC 404 reduced and telluric-corrected data cubes from the Jn2 band (left) and the Kbb band (right) collapsed along the wavelength axis.  The displayed maps were made by taking the average value of the spectrum in each spatial pixel (spaxel) across the entire wavelength range in each band and are thus dominated by that band's continuum emission.  The origins of these and all subsequent plots are set to the central peak of the Jn2/Kbb continuum emission as determined by a two dimensional Gaussian fit to the emission in each collapsed data cube (also indicated by the gray cross hairs).  We find that the nuclear J and K band continuum emission is compact and point like.
}
\end{figure}

\begin{figure}[th]
\begin{center}
 	\includegraphics[width=10cm]{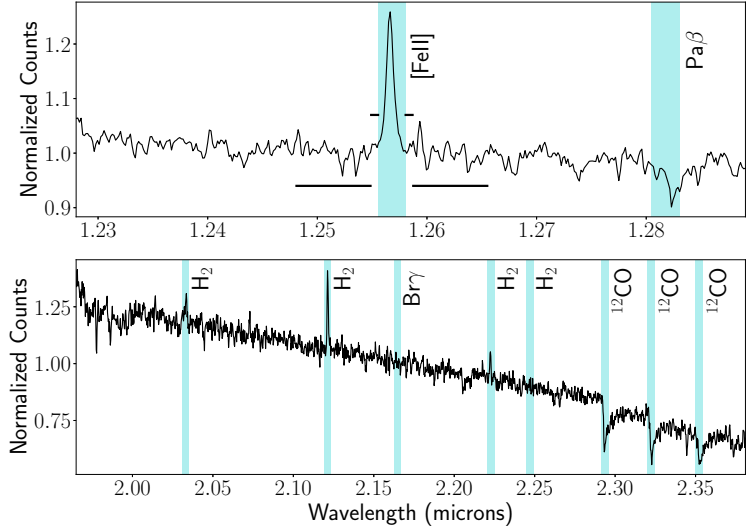}	
\end{center}
\caption
{ \label{fig:examplespectrum} Spectrum of the nucleus of NGC 404 in the Jn2 (top) and Kbb (bottom) bands.  The vertical bars indicate the wavelengths of the various spectral features at the systemic velocity of NGC 404 (-48 km s$^{-1}$) and span a $\pm$300 km s$^{-1}$ velocity range around the central wavelength.  
Horizontal lines indicate the spectral ranges used to determine the continuum (shown above the spectrum) and noise (shown below) for the [FeII] line (see Sec.~\ref{sec:linesnr}).  
Each spectrum was extracted from the Jn2/Kbb data cubes using an aperture centered on the continuum peak with a radius of 5 spaxels or 2.6 pc.  
The spectra show emission from the [FeII] line in the Jn2 band and multiple H$_2$ lines in the Kbb band.  Stellar absorption features from Pa$\beta$ and the $^{12}$CO bandheads are also visible.  Although not detected in this central aperture, Pa$\beta$ and Br$\gamma$ are also seen in emission to the north of the NGC 404 nucleus (see Fig.~\ref{fig:linemaps}).  
}
\end{figure}

Fig.~\ref{fig:reducedcubes} shows the final Jn2 and Kbb 35 mas data cubes of NGC 404 averaged across all wavelengths in the filter and Fig.~\ref{fig:examplespectrum} shows an example spectrum extracted from each of these data cubes.  The peaks of the Jn2 and Kbb continuum emission as determined by a two dimensional Gaussian fit to the spectrally-averaged data cubes were used to spatially register the two data sets to each other.  We find that the continuum emission is point-like in the OSIRIS data and has an average FWHM of 0.16 arcsec and 0.13 arcsec respectively; however, it is difficult to determine whether it is a true point source or slightly resolved because the core of the galaxy was used as the tip/tilt star.  Since the 35 mas data from 8 October 2015 are deeper and have a finer spatial sampling, analyses derived from these data are presented in subsequent sections of this work unless otherwise noted.

The instrumental dispersion and its variation across the mosaicked field of view were estimated by measuring the widths of OH sky lines in different apertures.  A set of 4 apertures were chosen for each filter to fall in sky regions separated from the continuum emission of the galaxy nucleus.  In the Jn2 band, the 4 apertures are offset from the continuum center by $\pm$0.5 arcsec in the x (east/west) direction and $\pm$0.67 arcsec in the y (north/south) direction.  In the Kbb band, the 4 aperture locations have the same x value as the continuum center and are offset in the y direction by $\pm$1.33 and $\pm$0.67 arcsec.  
The OH line widths in each aperture were estimated by Gaussian fits to 3 emission lines.  In the Jn2 band the chosen OH lines are close in wavelength, so their line profiles are fit simultaneously with a model consisting of a sum of 3 Gaussians.  The continuum in these spectra was estimated separately by performing a parabolic fit to the majority of the spectrum, with the exact wavelength range adjusted to give a good estimate to the wavelength range of the OH lines on the red side of the bandpass.  In the Kbb band, the OH lines are well separated, so the line is fit individually and the continuum was modeled by adding a delta offset to the Gaussian line profile model.  
The median values of the Jn2 and Kbb sigma values for the OH lines are 0.21 and 0.25 nm respectively.  
These instrumental line widths are equivalent to velocity dispersions of 50 km s$^{-1}$ in the Jn2 band and 35 km s$^{-1}$ in the Kbb band.

\section{Data Analysis and Results}
\label{sec:analysis}

\subsection{Line Emission S/N Estimate}
\label{sec:linesnr}

S/N ratios for the [FeII] and 1-0 S(1) H$_2$ emission lines were computed for every spaxel in the mosaicked data cubes for use in subsequent analysis steps.   
The continuum for this calculation was taken as the average of counts in the wavelength channels from -450 to -300 km s$^{-1}$ and from +300 to +450 km s$^{-1}$ relative to the systemic velocity.  The noise {was} estimated by taking the standard deviation of 10 wavelength channels on either side of the line, starting at $\pm$300 km s$^{-1}$ and moving outward.  Every fifth wavelength channel {was} used for the noise estimate to avoid the correlated noise present in adjacent wavelength channels.  

Signal and noise maps {were} created for the [FeII] and 1-0 S(1) H$_2$ emission lines by computing the signal and noise for each emission line in each spaxel.  Due to the noise present in the signal and noise estimates, the signal and noise maps {were} then smoothed using a 2D gaussian with a sigma of 3.5 pixels in both the x and y directions.  This smoothing {was} implemented using the \texttt{gaussian\_filter} function in the \texttt{ndimage} module of the SciPy library, a scientific computing library for the Python programming language.  The ratio of these smoothed maps {was} then used as the spatial map of the S/N ratios for each of the [FeII] and 1-0 S(1) H$_2$ emission lines in subsequent analysis steps.   The values of the smoothed S/N maps range from 0 in locations where there is no detected line flux to 25 in the spaxels with the maximum line flux. 

\subsection{Velocity-Integrated Emission Line Flux Maps}
\label{sec:linemaps}

Spatial maps of the continuum-subtracted emission line fluxes of the [FeII], Pa$\beta$, 1-0 S(1) H$_2$, and Br$\gamma$ lines {were} constructed to compare the morphology of the different lines.  The emission line flux in each individual spaxel {was} found by summing the continuum-subtracted counts in 7 wavelength channels (a velocity range of approximately $\pm$100 km s$^{-1}$) centered on the emission line wavelength.  This velocity range was chosen to capture the majority of the line emission while still excluding channels without emission that would only add noise to the line flux measurement.  For the [FeII] and 1-0 S(1) H$_2$ emission lines,  velocity information {was} used to determine the emission line wavelength in spectra with an emission line S/N greater than 10 (see Sec.~\ref{sec:linesnr}).  For spaxels with line S/N less than 10, the emission line wavelength {was} set as the vacuum wavelength of the line shifted to the systemic velocity of the galaxy.  
For the Pa$\beta$ and Br$\gamma$ emission lines, the systemic velocity of the galaxy {was} used to determine the emission line wavelength in every spaxel, since these lines are detected with high significance in only a few spatial locations.  
The continuum around each line {was} estimated using the average of counts in the wavelength channels from $-450$ to $-300$ km s$^{-1}$ and from +300 to +450 km s$^{-1}$ relative to the systemic velocity of the galaxy. 

\begin{figure}[t]
	\begin{center}
    \begin{tabular}{c}
	\includegraphics[width=11.0cm]{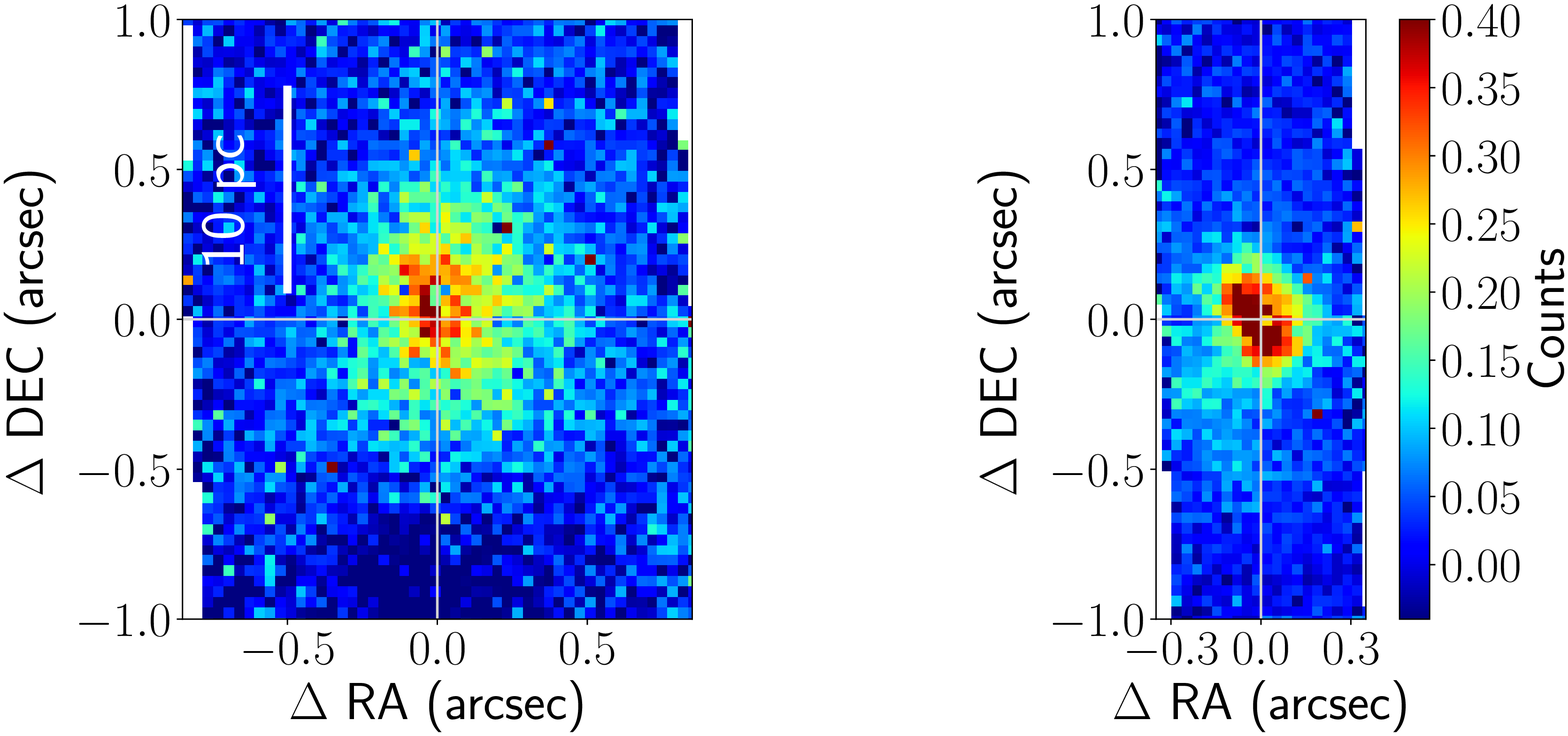} \\
    \includegraphics[width=11.0cm]{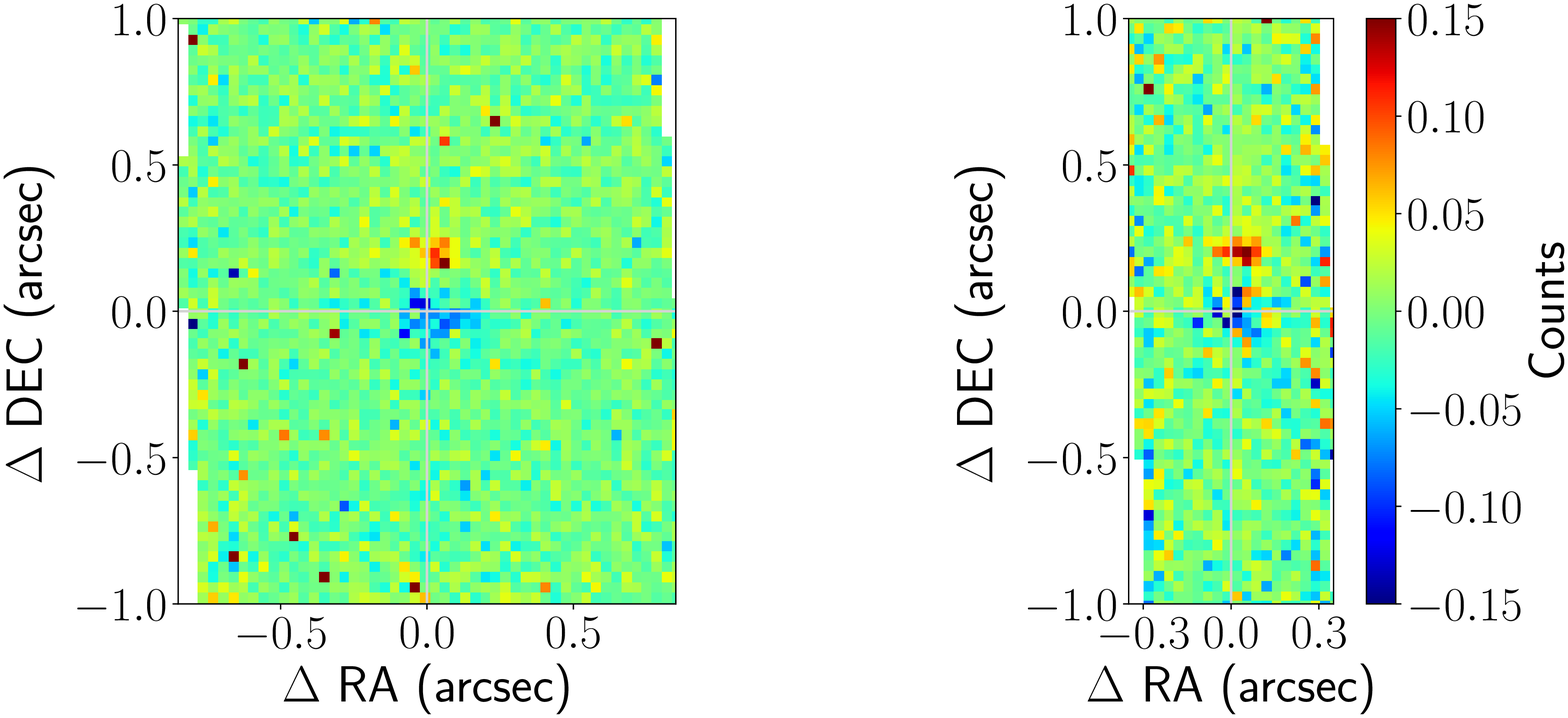}    
    \end{tabular}
	\end{center}
\caption
{ \label{fig:linemaps} Velocity-integrated emission line flux maps across the central 20 pc of NGC 404 for four emission lines: [FeII] (top left), 1-0 S(1) H$_2$ (top right), Pa$\beta$ (bottom left), and Br$\gamma$ (bottom right).  The [FeII] and Pa$\beta$ maps {were} derived from the Jn2 data and the H$_2$ and Br$\gamma$ maps are derived from the Kbb data.  These line maps are the sum of the continuum-subtracted emission in wavelength channels approximately $\pm100$ km s$^{-1}$ around the line relative to the systemic velocity of NGC 404.  The displayed maps are subsets of the full field of view of the mosaicked data cubes shown in Fig.~\ref{fig:reducedcubes} (1.65 x 3.29 arcsec in the Jn2 band and 0.70 x 3.33 in the Kbb band) and show the regions in which line emission is detected.  
The [FeII] emission is diffuse and extended out to $\sim$0.5 arcsec (8 pc) from the nucleus of NGC 404.  The bulk of the H$_2$ emission is more compact, with additional regions of emission to the bottom left of the nucleus.  The Pa$\beta$ and Br$\gamma$ lines are seen in absorption at the nucleus (in blue) and in emission $\sim$0.2 arcsec (3 pc) north of the nucleus (in red).  
}
\end{figure}

The resulting emission line flux maps for 4 emission lines are shown in Fig.~\ref{fig:linemaps}.  We find that the morphologies of the [FeII], H$_2$, and Pa$\beta$/Br$\gamma$ lines vary at the pc spatial scales probed by these data.  The [FeII] emission is diffuse and extended out to $\sim$0.5 arcsec (8 pc) from the nucleus of NGC 404, while the H$_2$ is more spatially compact.  The majority of the H$_2$ emission is contained in a central peak coincident with the continuum center of the K band emission, with weaker H$_2$ emission visible to the southeast of the nucleus.
Pa$\beta$ and Br$\gamma$ are seen both in absorption at the nucleus and in emission $\sim$0.2 arcsec (3 pc) north of the nucleus of NGC 404.

\subsection{Velocity-Resolved Emission Line Flux Maps}
\label{sec:velchannelmaps}

The emission morphology of each line can also be viewed as a function of velocity to reveal morphological structures that only appear at specific velocities.   
First, continuum-subtracted line cubes for each emission line {were} created by selecting a subset of wavelength channels around the emission line and then estimating and subtracting a continuum value for every spatial pixel.  The continuum value {was} determined using the same method as for the velocity-integrated line maps, by taking the average of the counts in wavelength channels from $-450$ to $-300$ km s$^{-1}$ and from +300 to +450 km s$^{-1}$.  
The wavelength ranges of the line cubes {were chosen to} span at least $\pm$300 km s$^{-1}$ around the lines relative to the systemic velocity of the galaxy to fully cover the wings of each emission line.  The wavelength channels of each line cube {were} then resampled to velocities from -250 to 250 km s$^{-1}$ with a spacing of 25 km s$^{-1}$.  This resampled spacing is slightly smaller compared to the native wavelength channel spacing of the OSIRIS data cubes ($\sim$35 km s$^{-1}$ in both Jn2 and Kbb bands).
The final emission line flux maps at each velocity {were} then made by taking the average of the channel at the given velocity and the two channels on either side ($\pm$25 km s$^{-1}$) in the resampled line cubes.  

\begin{sidewaysfigure}[th]
\begin{center}
\includegraphics[width=21cm]{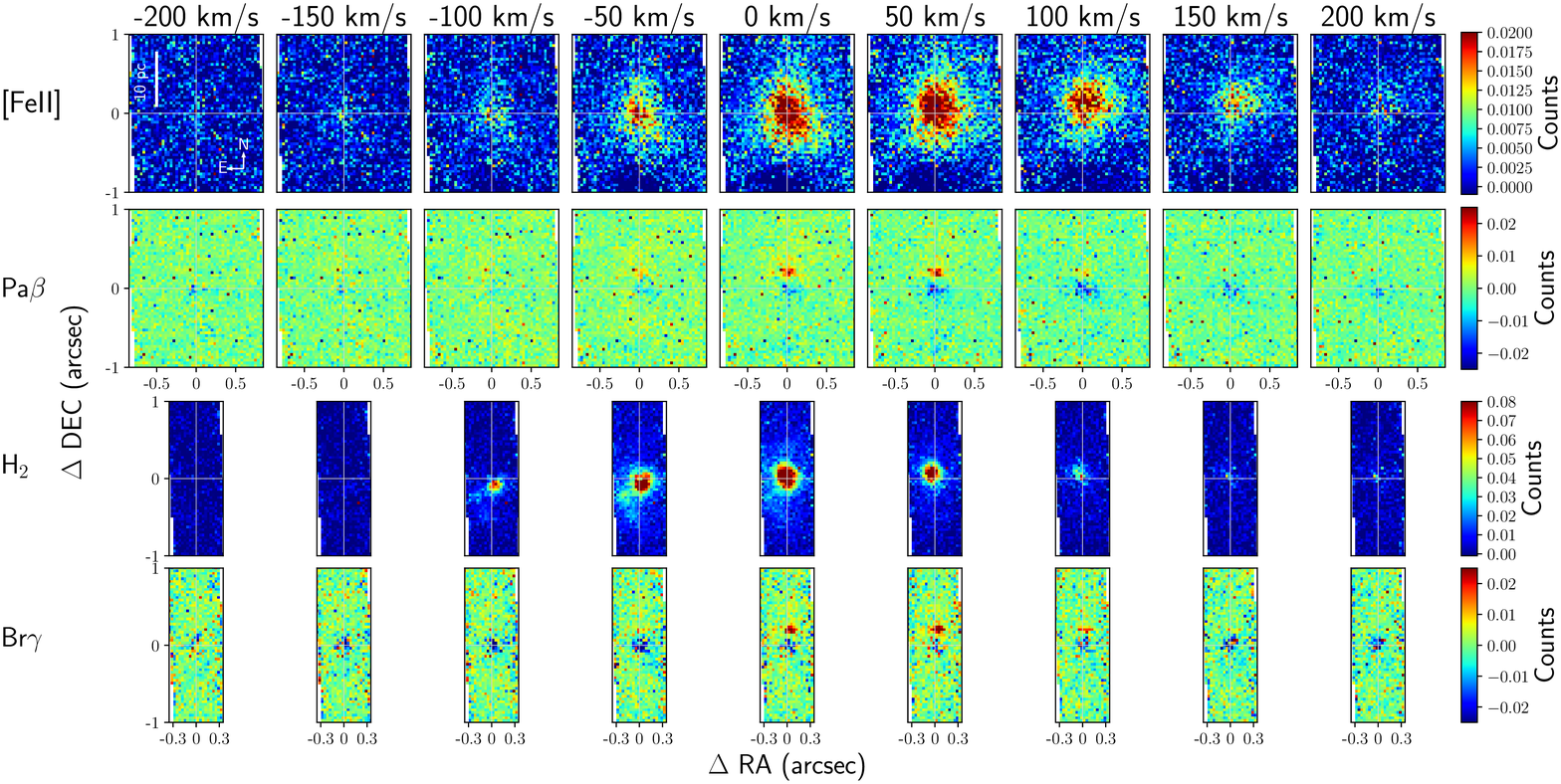}
\end{center}
\caption
{ \label{fig:velchannelmaps} Velocity-resolved emission line maps for emission lines: [FeII], Pa$\beta$, 1-0 S(1) H$_2$, and Br$\gamma$.  The displayed velocity is relative to the systemic velocity of the galaxy.  The cross hairs indicate the continuum center of the galaxy in the Jn2 ([FeII], Pa$\beta$) or Kbb (H$_2$, Br$\gamma$) bands.  These maps show that the [FeII] emission is extended and diffuse compared to the compact H$_2$ and Pa$\beta$/Br$\gamma$ emission and spans the largest velocity range of all the emission lines (-150 to +200 km s$^{-1}$).  At -50 and 0 km s$^{-1}$, the [FeII] emission is extended to the north and to the southwest of the nucleus, coincident with the arc-like structure seen in the 50 mas data (see Fig.~\ref{fig:feII_hst_compare}).  The central peak of the H$_2$ emission shifts position from the southwest of the continuum center to the northeast with increasing velocity, a signature of rotational motion.  Two additional blue-shifted knots of H$_2$ emission to the southeast of the nucleus are visible in the -50 and 0 km s$^{-1}$ maps.
}
\end{sidewaysfigure} 

Fig.~\ref{fig:velchannelmaps} shows the resulting velocity-resolved line flux maps for the [FeII], Pa$\beta$, 1-0 S(1) H$_2$, and Br$\gamma$ emission lines. The [FeII] emission shows a complex velocity structure, with the brightest and most extended emission at 0 and 50 km s$^{-1}$.  Low surface brightness [FeII] emission extends out to high velocities from -150 to +200 km s$^{-1}$.  
At -50 and 0 km s$^{-1}$ the shape of the [FeII] emission extends to the north and southwest of the nucleus.  This extended structure is coincident with an arc-like feature visible at these velocities in the 50 mas data set shown in Fig.~\ref{fig:feII_hst_compare}. 
The H$_2$ emission is more spatially compact than the [FeII] emission and spans a smaller overall velocity range of $\pm$100 km s$^{-1}$.  The emission peaks in the 0 km s$^{-1}$ channel where it aligns with the continuum center of the Kbb emission.  The $\pm$50 km s$^{-1}$ H$_2$ emission is offset from the continuum center to the southwest and northeast respectively, showing that two sides of the central molecular gas emission have different velocities, which is similar to what is expected from a rotating disk with a position angle {of its rotation axis equal to} about $-45^\circ$ on the plane of the sky.  
The two knots of H$_2$ emission offset from the continuum center to the southeast are brightest in the -50 km s$^{-1}$ map, revealing their blue-shifted velocity offset from the central region of molecular emission.  These blue-shifted H$_2$ structures are not matched in the [FeII], Pa$\beta$, or Br$\gamma$ emission lines.  The Pa$\beta$ and Br$\gamma$ maps show similar morphologies to each other across all velocity channels.  The knot of emission north of the nucleus seen in these lines spans from -50 to 100 km s$^{-1}$, with the brightest emission found at 0 and 50 km s$^{-1}$.  The velocities of the brightest Pa$\beta$/Br$\gamma$ emission match that of the brightest [FeII] emission.  Additionally, this bright emission region is spatially located where the [FeII] emission is also very strong.  Pa$\beta$ and Br$\gamma$ are also seen in absorption at the center of the NGC 404 nucleus, which can be used to constrain the stellar population in this central region.

\subsection{Emission Line Kinematics: [FeII] and H$_2$}
\label{sec:linevelmaps}
The velocity shifts of [FeII] and the 1-0 S(1) H$_2$ emission lines in a given spectrum {were} estimated using a Gaussian fit to the line profiles. The fits {were} performed in a wavelength range of $\sim$0.003 $\mu$m around the vacuum wavelength of the emission lines.  This wavelength range corresponds to approximately $\pm$350 km s$^{-1}$ in the Jn2 band and $\pm$200 km s$^{-1}$ Kbb band.  The model parameters include the central wavelength, the sigma, and the integrated flux of the Gaussian emission line as well as a vertical offset to model the continuum in the region around the line.  The central wavelength and the sigma of the best-fit Gaussian {were} converted into the velocity and the velocity dispersion of the emission line.  These fits {were} performed on each individual spectrum and on spatially binned spectra in square bins that are 2 spaxels (70 mas) on a side.  These binned spectra {were} created by taking the median of the 4 individual spectra in each bin.  

\begin{figure}[th]
\begin{center}
\begin{tabular}{c}
	\includegraphics[width=11cm]{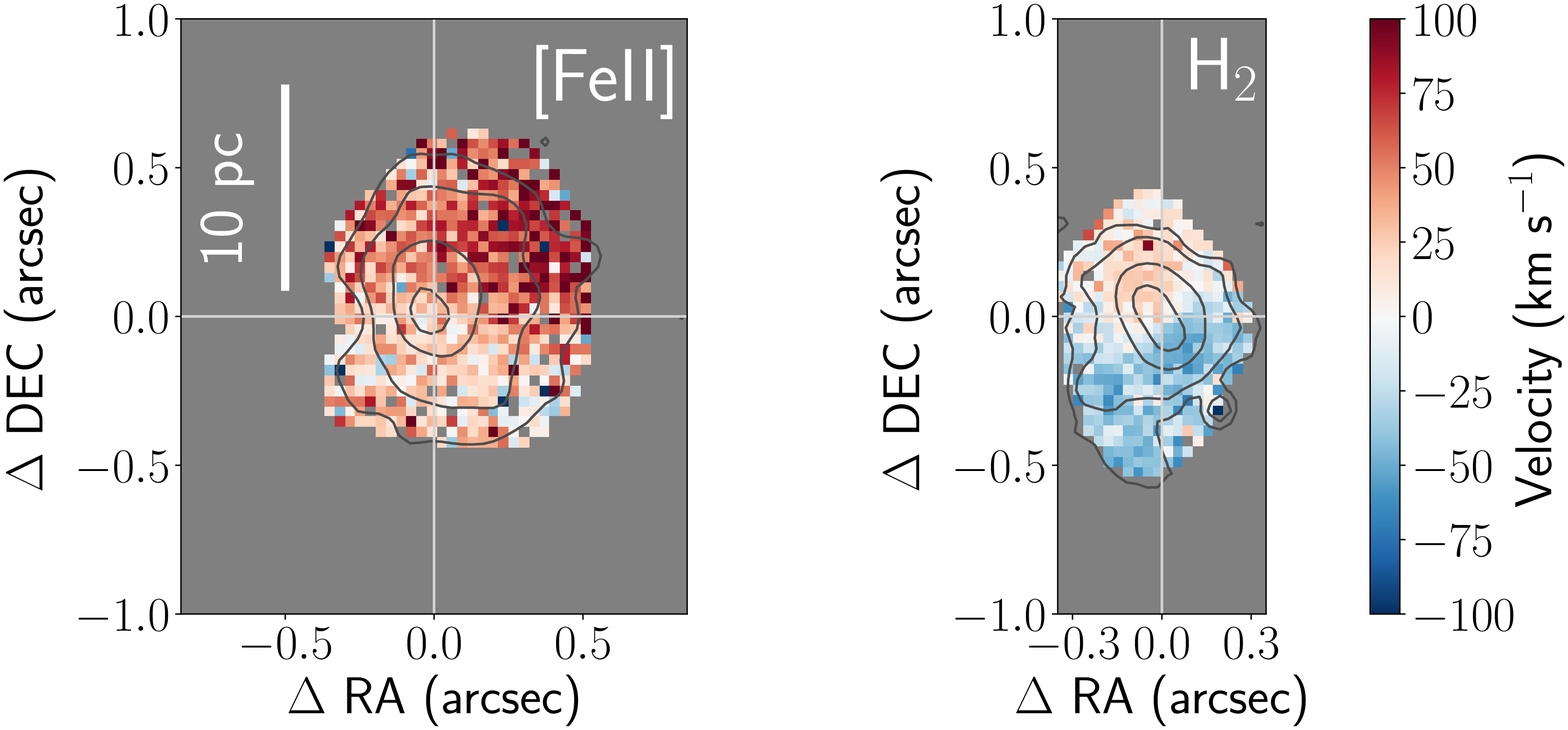} \\
    \includegraphics[width=11cm]{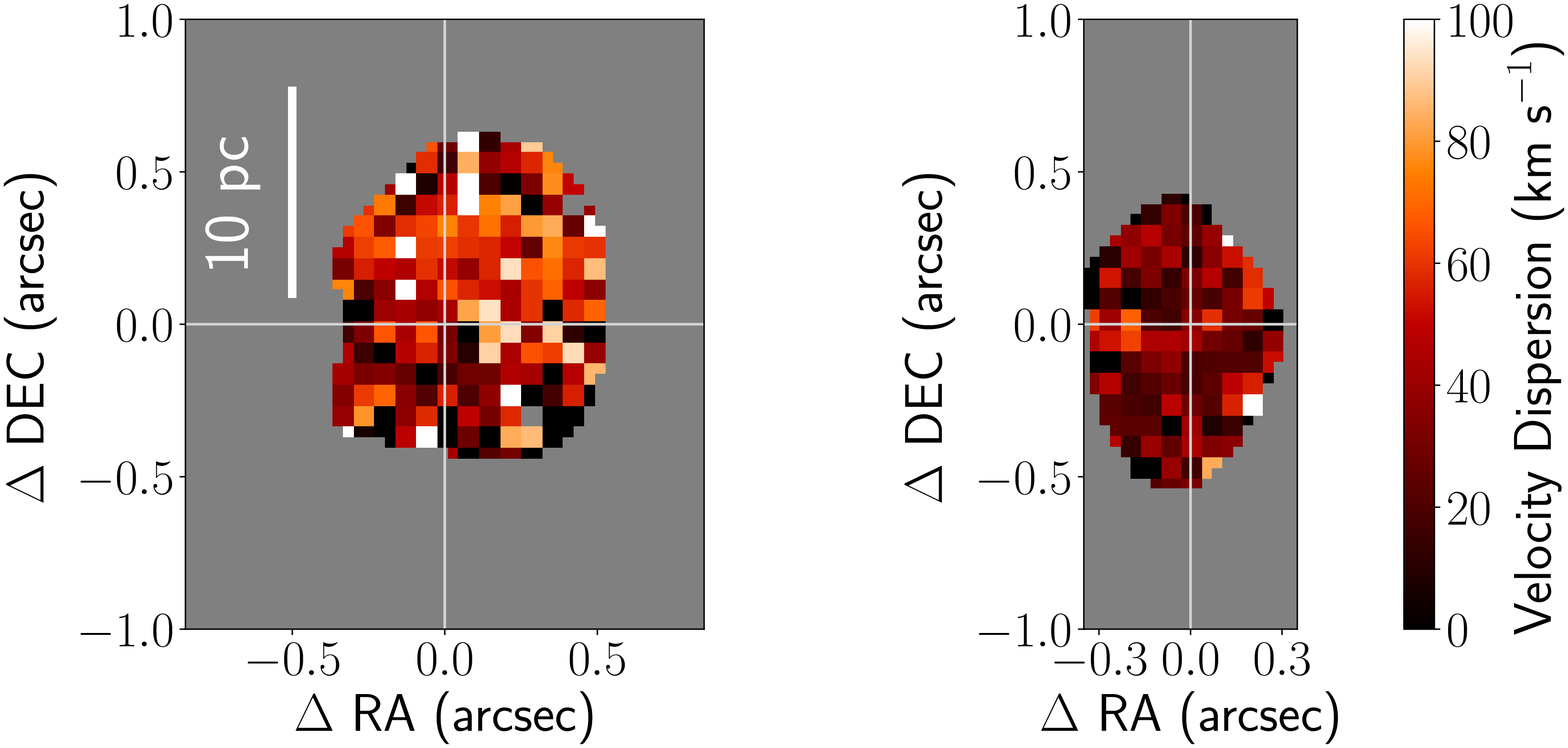}
\end{tabular}
\end{center}
\caption
{ \label{fig:veldispmaps} Velocity (top row) and velocity dispersion (bottom row) for the [FeII] (left column) and 1-0 S(1) H$_2$ (right column) emission lines.  The velocity maps are shown for every individual spaxel and the velocity dispersion maps are shown for spectra in spatial bins measuring 2 by 2 spaxels (70 mas).  Contours showing the morphology of the [FeII] and 1-0 S(1) H$_2$ line emission as displayed in Fig.~\ref{fig:linemaps} are overplotted in dark gray.  The velocities are shown relative to the systemic velocity of NGC 404 and with the red colors indicating red shifts and the blue colors indicating blue shifts.  The instrumental dispersions {were} subtracted from the measured velocity dispersions to yield the displayed dispersion {maps}.  Individual spaxels with a line S/N of less than 10 are masked.  Spaxels whose Gaussian line fit did not converge or whose best fit velocity falls outside the fitting range are also masked.  We find that the [FeII] and H$_2$ show different kinematics with velocity gradients that are roughly perpendicular to each other.  The [FeII] line reaches high velocities up to 80 km s$^{-1}$ and has a higher dispersion than the molecular gas.  The velocities of the central H$_2$ emission are dominated by rotational motion with a velocity gradient of $\pm$30 km s$^{-1}$.  The two knots of H$_2$ emission to the southeast of the nucleus are blue shifted by $\sim$30 km s$^{-1}$.
}
\end{figure}

The velocity and velocity dispersion maps for the [FeII] and 1-0 S(1) H$_2$ emission lines are displayed in Fig.~\ref{fig:veldispmaps}.  The velocity is {displayed} for each individual spaxel, while the velocity dispersion is {displayed for} the spatially binned spectra to reduce noise.  {To create the displayed} dispersion maps, the instrumental dispersion determined in Sec.~\ref{sec:observations} for each observing band {was} subtracted {in quadrature} from the measured velocity dispersion.  For the $\sim$5 bins with measured velocity dispersions less than the instrumental dispersion, the dispersion is set to 0 {in Fig.~\ref{fig:veldispmaps}} for display purposes.  
Individual spaxels with a S/N ratio less than 10 in the smoothed S/N maps are masked.  Additionally, spaxels whose Gaussian fit did not converge or whose best fit velocity falls outside the wavelength range are also masked.  The same mask {was} applied to both the velocity and the velocity dispersion maps.

The [FeII] and H$_2$ gases show very different velocity structures.  The H$_2$ velocity map is consistent with simple rotation centered on the nucleus of NGC 404 with a velocity shift of $\pm$30 km s$^{-1}$ across the central $\sim$15 pc of the galaxy.  The two peaks of H$_2$ emission offset to the southeast of the nucleus are found to be blue shifted by $\sim$30 km s$^{-1}$ relative to the systemic velocity of the galaxy.  The [FeII] velocity gradient is roughly perpendicular to the gradient of the molecular gas and spans a larger range of velocities, with the northwest region of emission having a velocities up to 80 km s$^{-1}$ relative to the systemic velocity of the galaxy.  This region with the highest [FeII] velocities also shows higher velocity dispersion with values up to 75 km s$^{-1}$.  The H$_2$ gas shows no spatial variation of the velocity dispersion and has generally lower dispersion than the [FeII] gas.

\subsection{Absorption Line Kinematics: CO bandhead}
\label{sec:COvelmap}

The CO bandhead is a set of strong absorption features primarily from the atmospheres of giant stars. 
With resolved spectroscopy this stellar absorption feature can be used to track the kinematics of the stars in the center of NGC 404.  We {measured} the velocity shift of the CO bandhead absorption feature in each individual spectrum relative to the spectrum of the spaxel that has the brightest Kbb continuum.  The velocity shift between a spectrum in a single spaxel and the reference spectrum {was} found by cross correlating the two spectra in the wavelength region containing the CO bandhead features (2.27 - 2.365 $\mu$m).  The peak of the resulting cross correlation {was} then found by fitting a Gaussian to the cross correlation function.  
Note that the stellar kinematics that we measure are therefore all relative to the velocity of this reference spectrum and are not absolute velocities like those we measure for the gas kinematics.

\begin{figure}[t]
\begin{center}
\includegraphics[width=6cm]{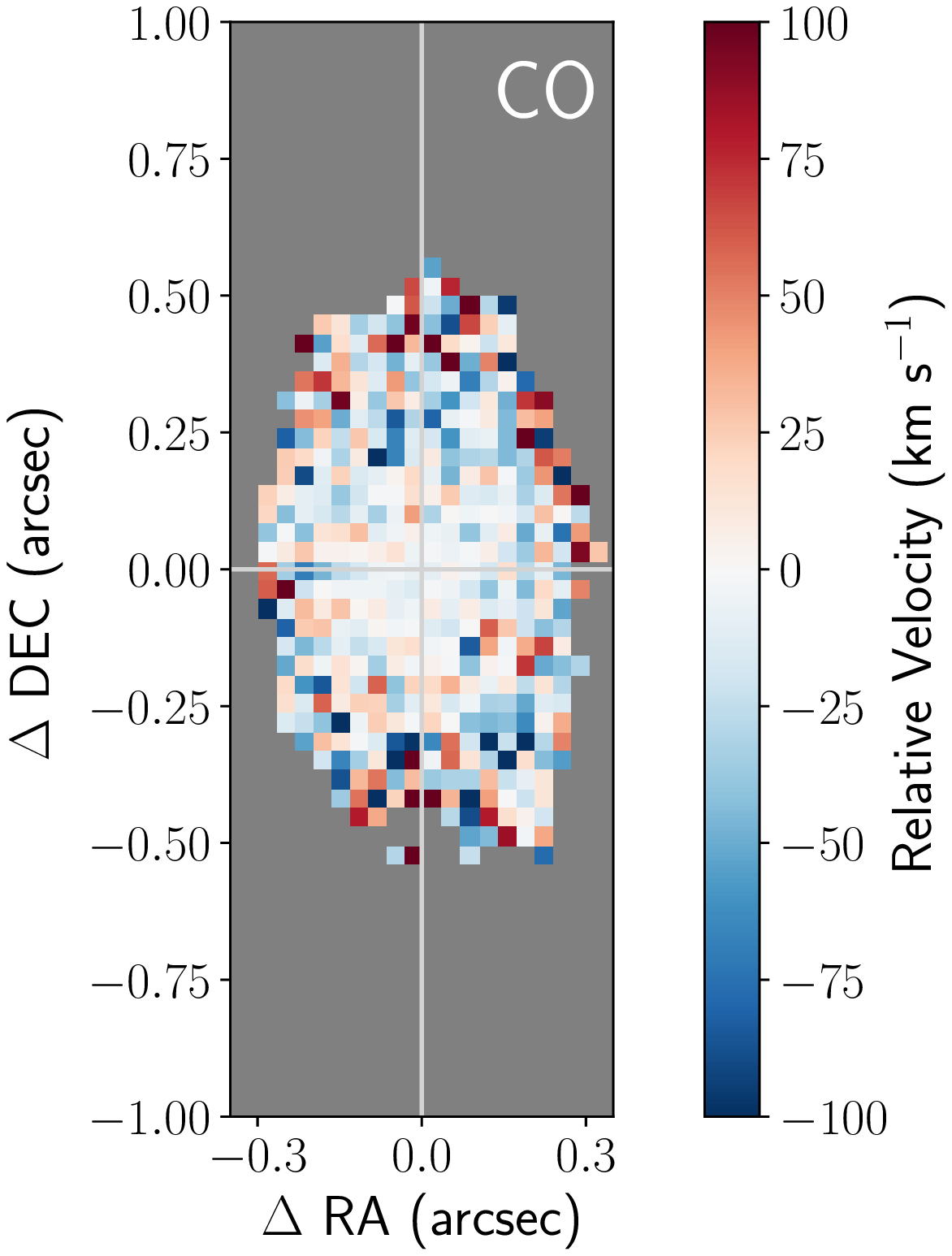}
\end{center}
\caption
{ \label{fig:COvelmap} Relative stellar velocity map as measured by the CO bandhead absorption feature.  The stellar velocities are displayed only for individual spaxels with continuum S/N greater than 7.0.  The velocity shifts {were} measured relative to the spaxel with the highest Kbb continuum emission, which is the spaxel at the origin of this plot.  The colorbar used here matches the colorbar used in Fig.~\ref{fig:veldispmaps} for the [FeII] and H$_2$ gas kinematics.  The CO velocity shows no significant variation across the central $\sim$15 pc of NGC 404 in contrast with the [FeII] and H$_2$ emission lines, thus revealing that the stars are not moving at the same velocities as the gas.  
}
\end{figure}

The resulting stellar velocity map is shown in Fig.~\ref{fig:COvelmap}.  Only spaxels with a continuum S/N greater than 7.0 are displayed.  This continuum S/N {was} computed using the DER\_SNR algorithm, which is a simple and robust algorithm that uses the entire spectral range to compute both the signal and the noise and accounts for correlated noise between neighboring wavelength channels \citep{2008ASPC..394..505S}.  
We find no significant variation in the stellar velocity within the central $\sim$15 pc of NGC 404.

\begin{figure}[t]
\begin{center}
\begin{tabular}{c}
	\includegraphics[width=11cm]{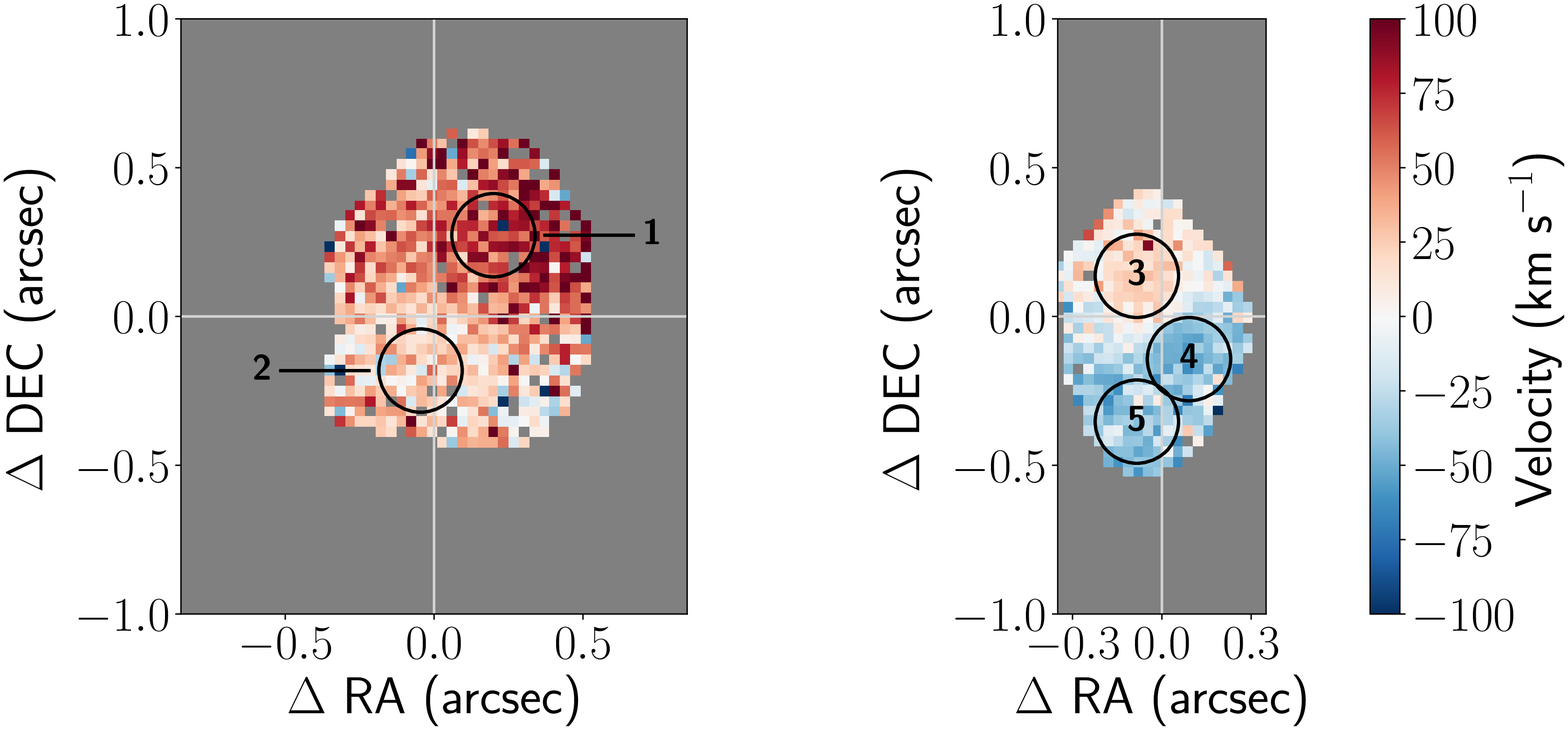} \\
    \includegraphics[width=11cm]{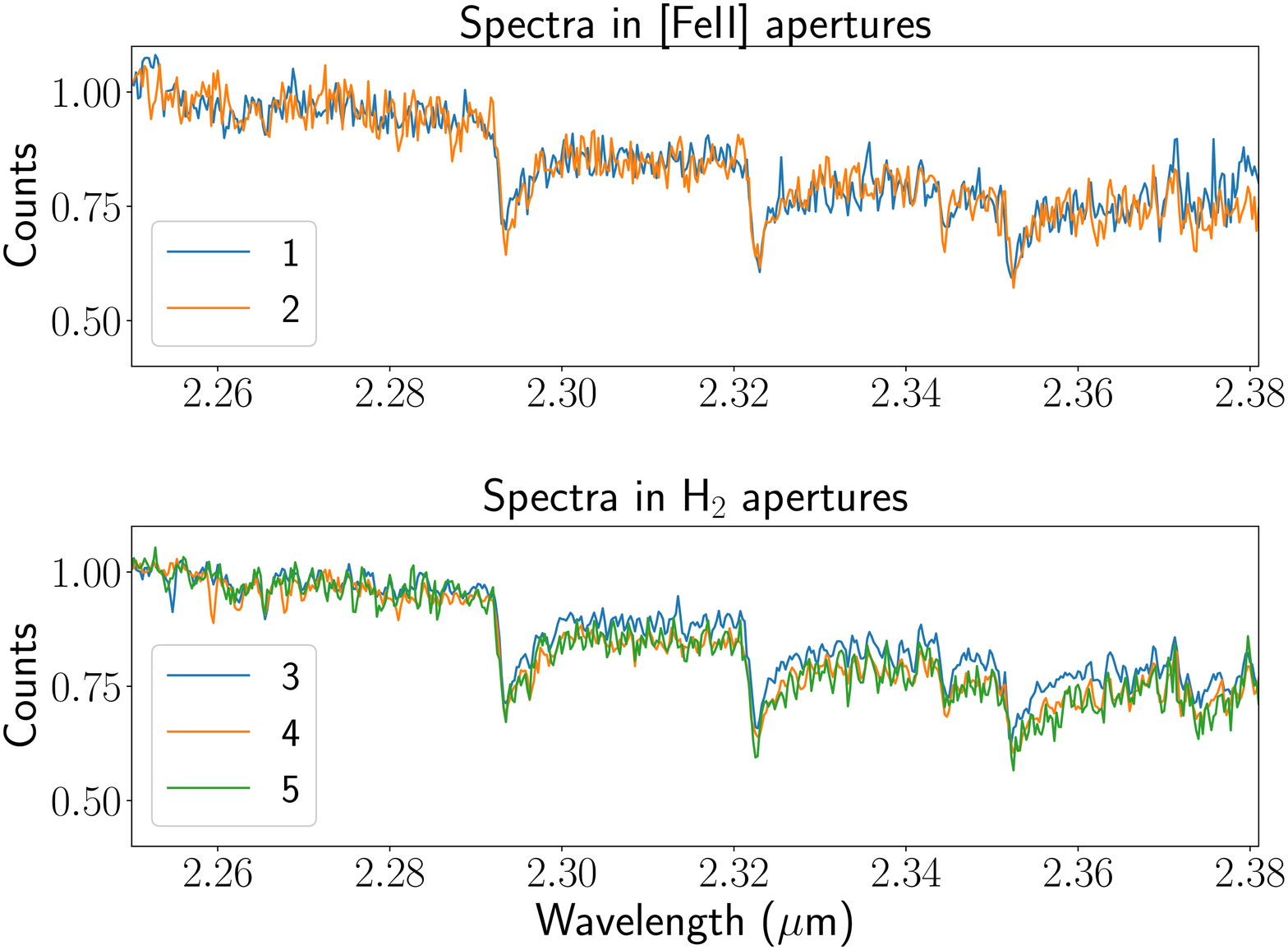}
\end{tabular}
\end{center}
\caption
{ \label{fig:CO_gas_velcomparison} 
Comparison of CO bandhead absorption features in 5 spatial apertures centered on regions of high/low [FeII] or 1-0 S(1) H$_2$ velocities, to confirm that the stellar velocity is constant even in regions with large gradients in ionized/molecular gas velocity.  The 5 circular apertures used are overplotted on the [FeII] and H$_2$ velocity maps in the top two panels.  The lower two panels display the 5 spectra extracted from these spatial apertures in the wavelength region around the CO bandhead spectral features.  The spectra are separated into two groups: the two apertures centered on regions of high and low [FeII] velocities and the three apertures centered on regions of high and low H$_2$ velocity.  No stellar velocity shift is visible within these two groups of spectra, indicating that the stars are not moving in the same velocity field as either the ionized or molecular gas.  Additionally, cross correlations measurements between spectra within these groups also give stellar velocity shifts consistent with 0 km s$^{-1}$.
}
\end{figure}

To confirm the lack of stellar velocity gradient in the data and to compare to the ionized and molecular gas velocities, we also {measured} the stellar velocity in apertures that have large differences in gas velocities.
Circular apertures with radii of 4 pixels (2 pc) {were} placed on 5 locations with high/low [FeII] or 1-0 S(1) H$_2$ velocities.  These apertures are overplotted on the [FeII] and 1-0 S(1) velocity maps and the spectra extracted from each aperture are shown in Fig.~\ref{fig:CO_gas_velcomparison}.  Pairs of CO spectra with high/low [FeII] or H$_2$ velocities {were} also cross correlated with each other to check for stellar velocity shifts. 
{We find that there is no stellar velocity shift visual in the extracted spectra and measure relative velocities of $-20$ $\pm$ 14 km s$^{-1}$, 0 $\pm$ 10 km s$^{-1}$, and 2 $\pm$ 12 km s$^{-1}$ from the spectra in apertures 1 and 2, 3 and 4, and 3 and 5 respectively. The errors were determined by the standard deviation of the velocity shifts measured from each of the 3 individual CO bandheads in the spectra. }  
{These stellar velocity shifts from the CO bandhead spectra are consistent with 0 km s$^{-1}$ and we therefore find that} the stellar velocity field is distinct from both the ionized and molecular gas kinematics.  

\subsection{Molecular Hydrogen Line Ratios} 
\label{sec:H2temps}

\begin{figure}[th]
\begin{center}
\begin{tabular}{cc}
\includegraphics[height=6cm]{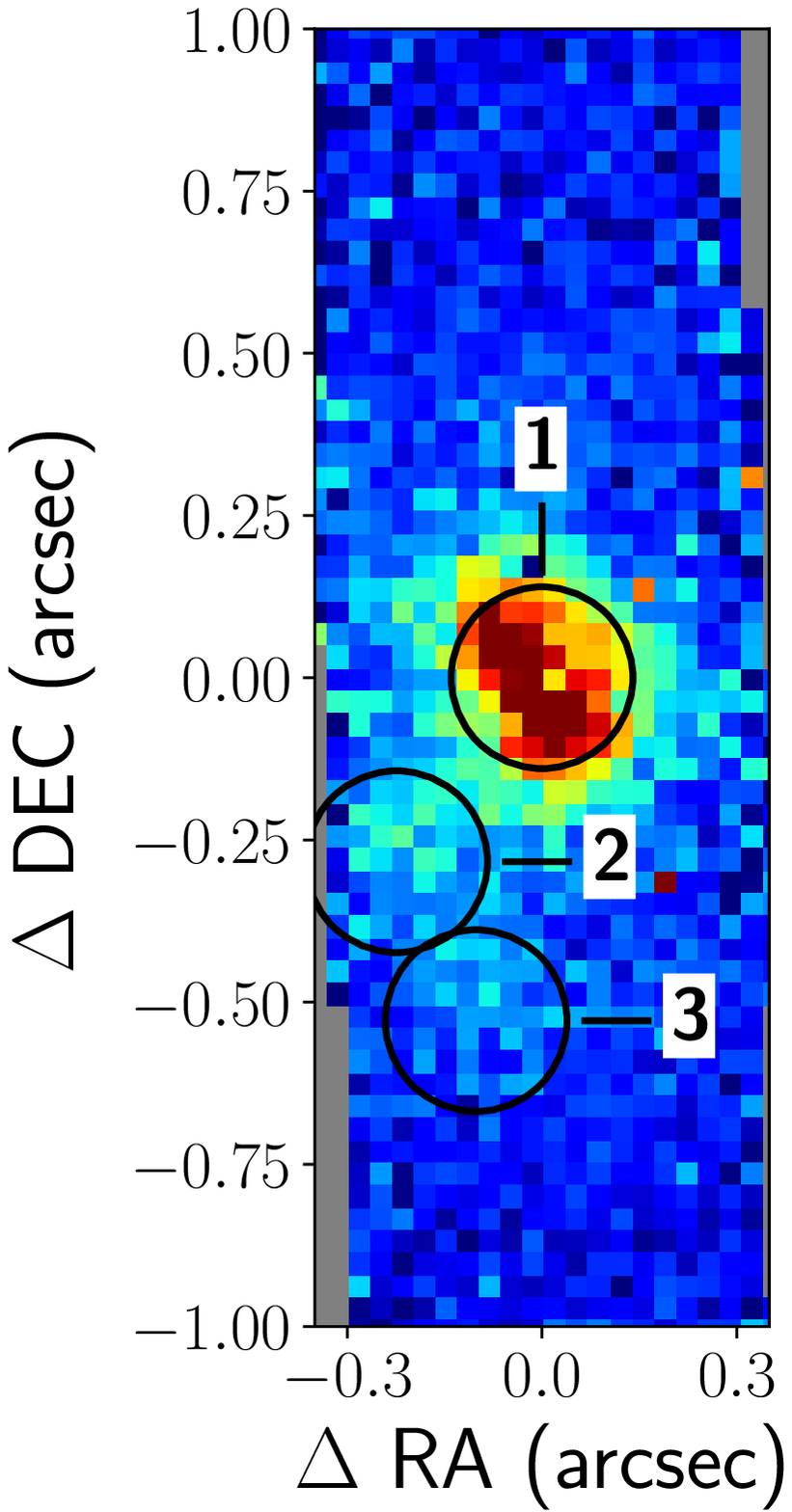} & \includegraphics[height=6cm]{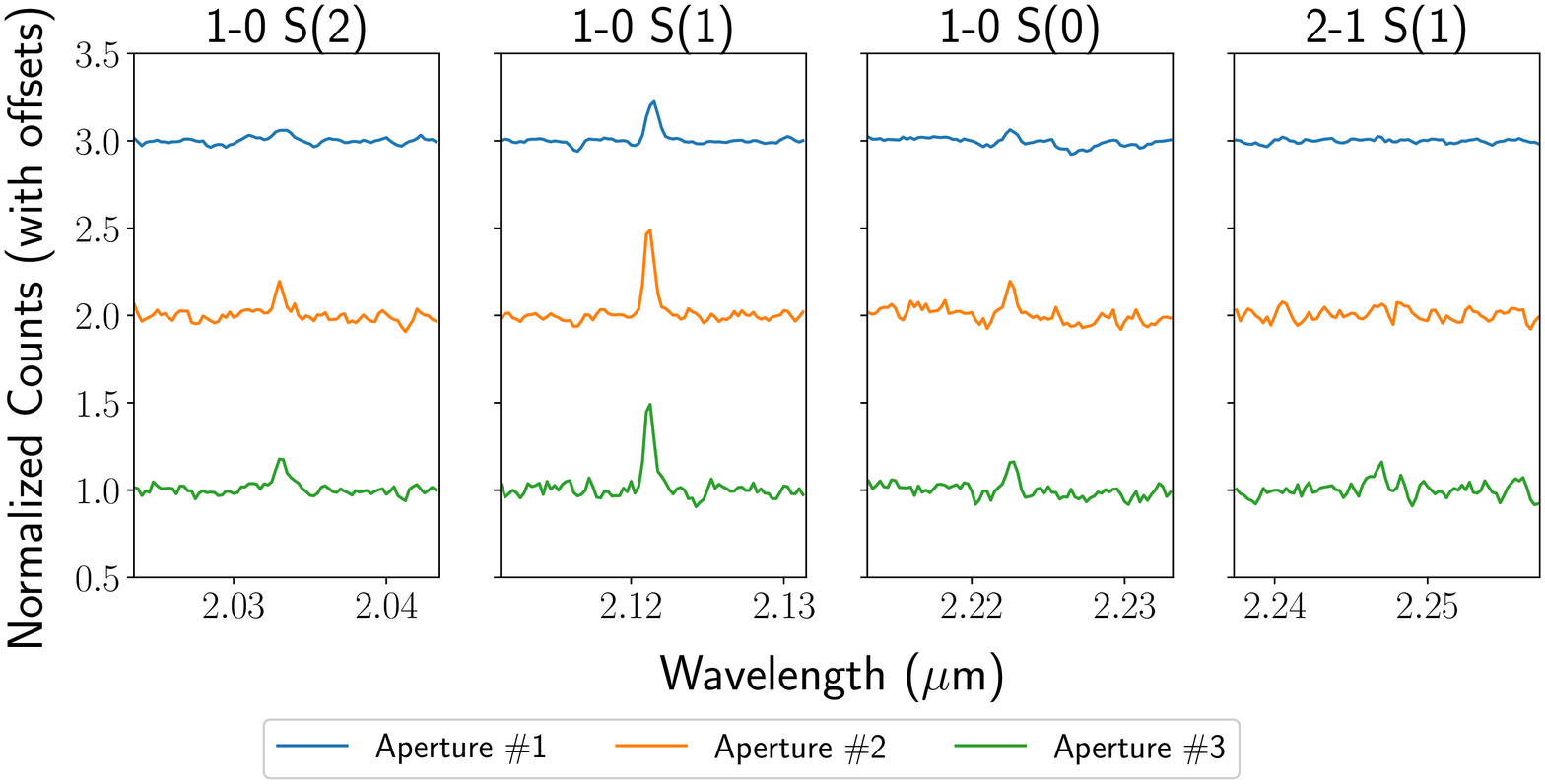} 
\end{tabular}
\end{center}
\caption
{ \label{fig:h2_aperture_spectra} Left: 1-0 S(1) H$_2$ emission line map with 3 apertures superimposed.  The apertures each have a radius of 4 pixels = 0.14 arcsec and {were} chosen to align with morphological features in the 1-0 S(1) emission map.  Right: Kbb spectra extracted from the three apertures in the wavelength regions around the detected H$_2$ emission lines.  These spectra {were} normalized to have a median of 1.0 and then an offset {was} applied for display purposes.  The wavelength regions around each H$_2$ line are plotted in separate panels.  The 1-0 S(2), 1-0 S(1), and the 1-0 S(0) H$_2$ lines are detected in all three apertures and the 2-1 S(1) line is only detected in the aperture \#3.}
\end{figure}

We {estimated} the flux of multiple molecular hydrogen lines within the wavelength range of the K broadband filter.  Combinations of these line fluxes can be used to estimate the rotational and vibrational excitation temperature of the H$_2$ and distinguish between different excitation mechanisms of this molecular gas.  In addition to the strongest 1-0 S(1) line of H$_2$ in the NGC 404 spectra discussed above, there are 3 detected H$_2$ lines in the Kbb bandpass: 1-0 S(0) ($\lambda_0$ = 2.2235 $\mu$m), 1-0 S(2) ($\lambda_0$ = 2.0338 $\mu$m), and 2-1 S(1) ($\lambda_0$ = 2.2477 $\mu$m).  We {estimated} these 4 line fluxes in 3 different circular apertures with radii of 4 pixels = 0.14 arcsec.  These apertures are centered on the different spatial structures observed in the 1-0 S(1) line map: the bright H$_2$ emission centered on the nucleus and the two weaker peaks of H$_2$ emission offset to the southeast of the nucleus. Fig.~\ref{fig:h2_aperture_spectra} shows the apertures superimposed on this line emission map and the corresponding extracted spectra.
The continuum {was} estimated globally in each of the three spectra using a line fit to a large subset of the wavelength range: 2.02 - 2.28 $\mu$m.  This range excludes regions on the blue side of the spectra that suffer from poor telluric correction and regions on the red side that contain the CO bandhead absorption features while still covering the wavelengths of all 4 H$_2$ lines.  

The H$_2$ line fluxes {were} then estimated from the continuum-subtracted spectra.  First, the overall molecular hydrogen velocity in each aperture {was} estimated using the 1-0 S(1) line, using the same Gaussian fitting method used for the individual spaxel velocity estimate (see Sec. \ref{sec:linevelmaps}).  The 1-0 S(1) velocities relative to the systemic velocity of NGC 404 in the apertures from north to south are $-5 \pm 3$, $39 \pm 2$, and $38 \pm 3$ km s$^{-1}$ respectively, where the error bars correspond to the fitting errors given by the curve\_fit algorithm from the optimize module of the SciPy library.  The flux of each line {was} then estimated by summing over wavelength channels with velocity in the range of $\pm$ 125 km s$^{-1}$ relative to the 1-0 S(1) velocity of each aperture.  
This velocity range was chosen using the 1-0 S(1) line in the central aperture by estimating its flux with a variety of velocity ranges and finding the minimum range that captured all of the line flux.

The errors on the line fluxes in each aperture spectrum {were} estimated by applying the line flux measurement method described above to areas of the spectrum with no line flux.  Wavelength values {were} sampled from a uniform distribution across the continuum fitting range of 2.02 - 2.28 $\mu$m.  Various spectral regions within this wavelength range {were} masked out, including around the 4 H$_2$ lines, the doublet absorption features from NaI (2.206/2.209 $\mu$m) and CaI (2.263/2.266 $\mu$m), and strong OH sky lines.  
The 675 wavelength values remaining after the mask {was} applied {were} then used as the central wavelength for line flux measurements.  The standard deviation of these resulting simulated line flux measurements {was} taken as the error on the measured H$_2$ line fluxes for each spectrum.  For H$_2$ lines with fluxes that are less than 3 sigma above the error, a 3-$\sigma$ upper limit on the flux {was} determined using the method described in Sec. \ref{sec:NIRdiagnosticratios}.  The velocity dispersion of the Gaussian line profile that sets the limiting flux {was} taken as 50 km s$^{-1}$ to match the measured dispersion of the 1-0 S(1) line, which is detected with high significance in all three apertures.

The line flux measurements and errors for each aperture are presented in Table \ref{tab:h2temps}.  This table also gives the H$_2$ line ratios 1-0 S(2)/1-0 S(0) and 2-1 S(1)/1-0 S(1) for each aperture. These line ratios are sensitive to the excitation mechanism of the molecular gas and are derived using fluxes from both ortho (odd $J$) or both para (even $J$) hydrogen molecules, so that a ortho/para ratio does not need to be independently assumed.
The 2-1 S(1) line {is} not detected above 3 sigma in apertures \#1 and \#2, so the upper limits on the line flux and the corresponding line ratio are reported.  The other emission lines were detected above 3 sigma in all apertures, and with the highest significance in aperture \#2, where the lower underlying continuum emission relative to the H$_2$ line emission leads to a lower line flux error.  We find that both the 1-0 S(2)/1-0 S(0) and 2-1 S(1)/1-0 S(1) line ratios are consistent among all three spatial locations.

{The H$_2$ line ratios measured here use pairs of emission lines that are far apart in wavelength and can therefore be affected by variations in Strehl ratio across the Kbb band.  
To quantify the affect of Strehl ratio variation across the H$_2$ line wavelengths, we divided the telluric-corrected Kbb data cube by a blackbody spectrum normalized to a median of 1.0 to remove the shape of NGC 404's continuum emission, so that any remaining variations in the aperture spectra are only due to variations in the Strehl ratio.  A temperature of 4925 K was chosen to match the F547M $-$ F814W (roughly V $-$ I) color of $\sim$0.95 measured in the unreddened region of the galaxy's nucleus using the Hubble Space Telescope 
(see Sec.~3.1 of \citet{2010Seth}; note that adjusting the temperature by $\pm$1000 K affects the blackbody spectrum shape by <1\% across the Kbb band). 
We compared the median value of the divided aperture spectra in regions $\pm$0.01 $\mu$m around the H$_2$ lines of interest and find that the Strehl ratio variations affect the line ratios at a level of 0.5 - 7\% across the two line ratios and the three apertures.  The 25 - 35\% statistical error on the line ratios reported in Table \ref{tab:h2temps} therefore dominates the error budget and the measured line ratios are not significantly affected by Strehl ratio variation in our data.}

\begin{deluxetable}{crrrrrrrrrr}
\tabletypesize{\scriptsize}
\tablewidth{0pt}
\tablecaption{H$_2$ Line Fluxes and Excitation Temperatures}
\tablehead{
  \colhead{Aperture} & 
  \multicolumn{4}{c}{Line Fluxes (Counts)} & 
  \colhead{Line Flux} &
  \colhead{1-0 S(2)/} &
  \colhead{2-1 S(1)/}  \\
  \colhead{Number} &
  \colhead{1-0 S(2)} & 
  \colhead{1-0 S(1)} & 
  \colhead{1-0 S(0)} & 
  \colhead{2-1 S(1)} & 
  \colhead{Error (Counts)} &
  \colhead{1-0 S(0)} &
  \colhead{1-0 S(1)} &  
}
\startdata
\#1  &  0.09 &  0.33 &  0.15 & $<$0.10 & 0.03 & 1.7 $\pm$ 0.6 & $<$0.30  \\ 
\#2  &  0.029 &  0.089 &  0.042 & $<$0.020 & 0.007& 1.5 $\pm$ 0.4& $<$0.23  \\
\#3  &  0.016 &  0.049 &  0.030 & 0.012 & 0.004 & 1.9 $\pm$ 0.5 & 0.25 $\pm$ 0.09
\enddata 
\label{tab:h2temps}

\end{deluxetable}

\subsection{[FeII]/Pa$\beta$ and H$_2$/Br$\gamma$ Line Ratios} 
\label{sec:NIRdiagnosticratios}

\begin{figure}[th]
\begin{center}
\begin{tabular}{c}
    	\includegraphics[width=12cm]{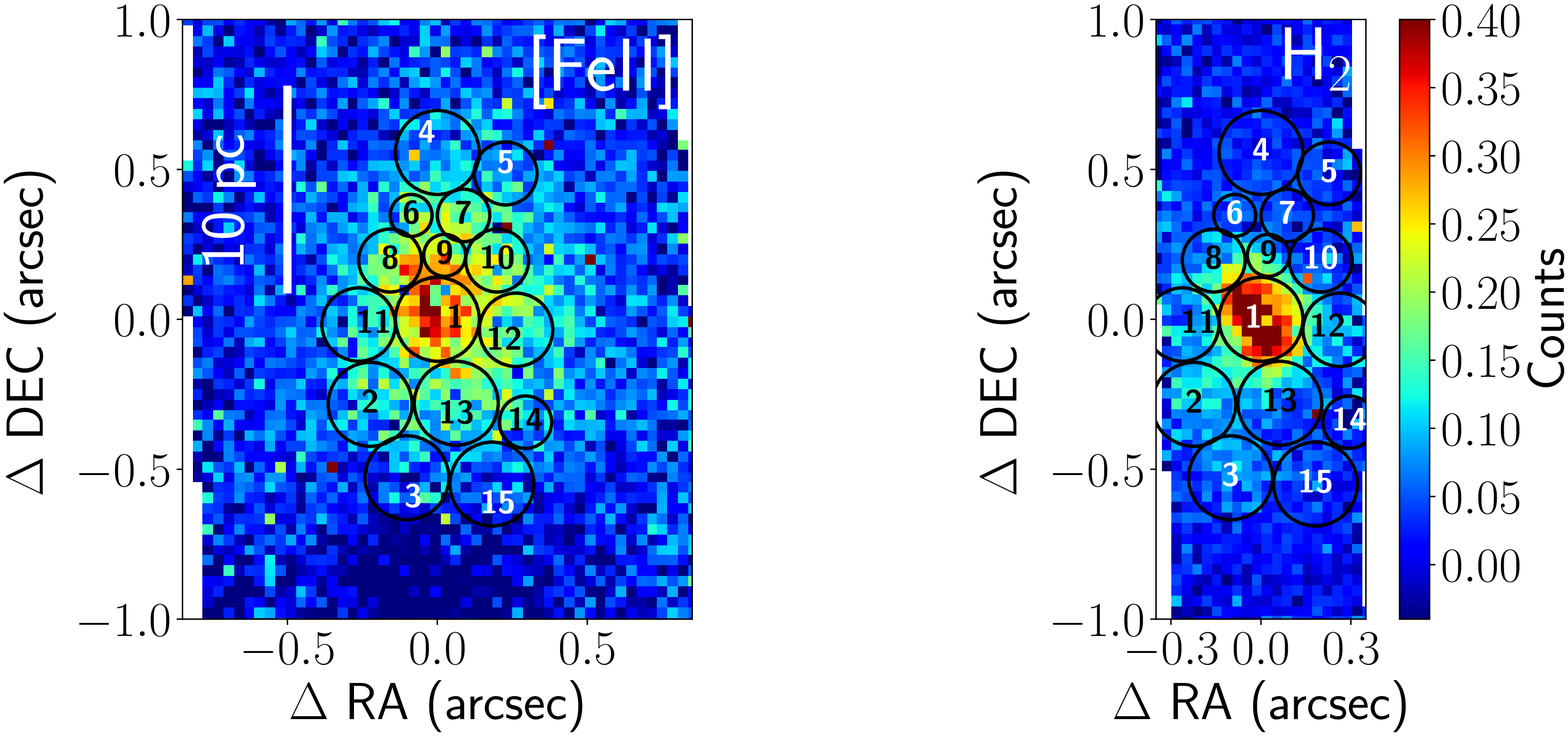} \\
    \includegraphics[width=12cm]{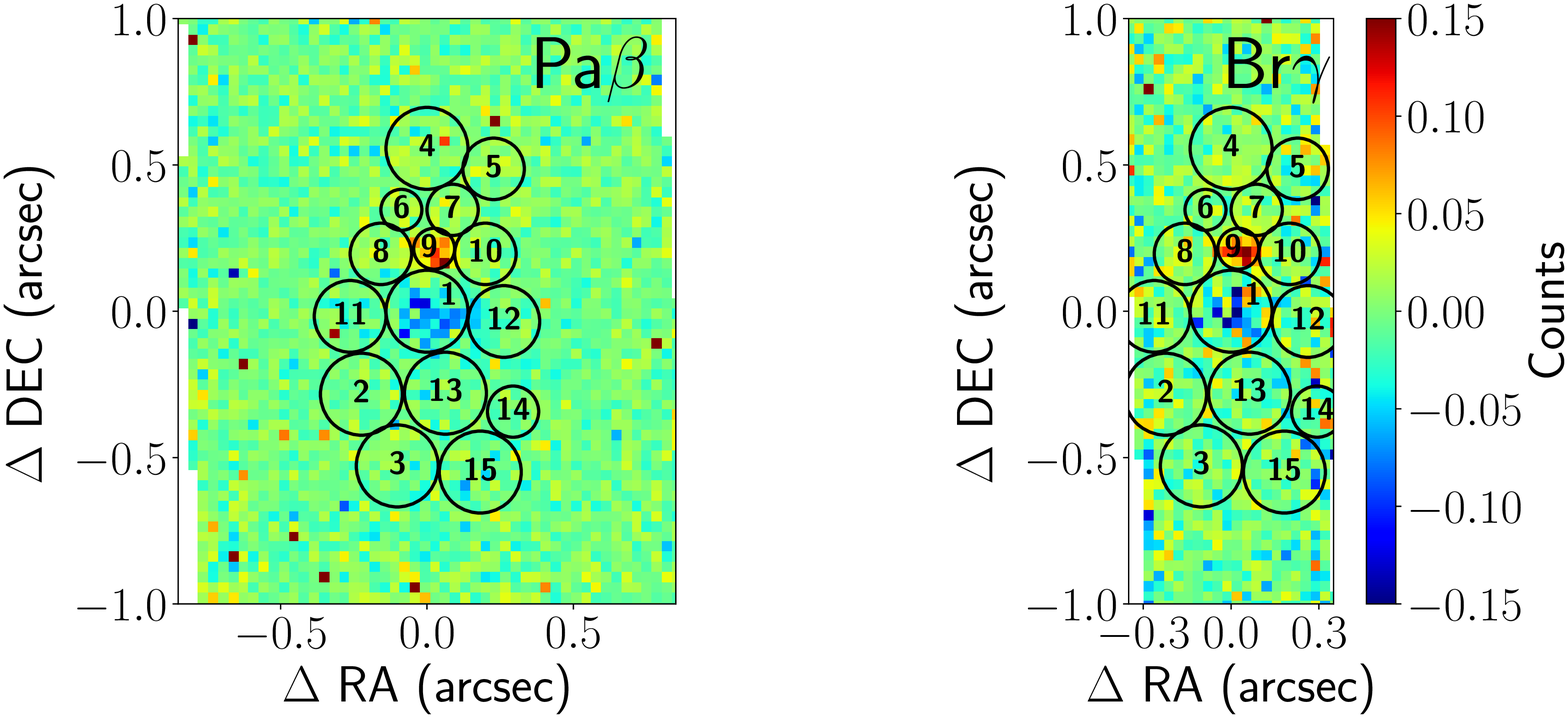}
\end{tabular}
\end{center}
\caption
{ \label{fig:lineratioaps} Circular apertures in which the [FeII]/Pa$\beta$ and 1-0 S(1) H$_2$/Br$\gamma$ are estimated overplotted on the [FeII], Pa$\beta$, 1-0 S(1) H$_2$, and Br$\gamma$ emission line maps from Fig. \ref{fig:linemaps}.  Apertures \#1, 2, and 3 are the same apertures used in the H$_2$ line ratio measurements described in Sec.~\ref{sec:H2temps}. Aperture \#9 was chosen to be centered on the peak of Pa$\beta$/Br$\gamma$ to the north of the NGC 404 nucleus.  Other apertures {were} used to fill in the remaining area of the [FeII] emission, the most spatially extended emission line. The radii of the apertures varies from 2 - 4 pixels (1 - 2 pc).}
\end{figure}

The line ratios of [FeII]/Pa$\beta$ and 1-0 S(1) H$_2$/Br$\gamma$ {were} measured to constrain the excitation mechanisms at different physical locations at the center of NGC 404. 
In general, the S/N of the hydrogen recombination lines is not high enough to estimate their line flux in any but a few spaxels (see Fig.~\ref{fig:linemaps}), so these line ratios {were} instead measured in various circular apertures placed within the shared Jn2/Kbb FOV with radii ranging from 2 - 4 pixels (1 - 2 pc).  
We {defined} these apertures by starting with the 3 apertures used for the H$_2$ line ratio measurements in Sec.~ \ref{sec:H2temps}, which are centered on the NGC 404 nucleus and two other areas of H$_2$ emission.  An additional aperture {was} applied to the location of the Pa$\beta$/Br$\gamma$ emission peak to the north of the nucleus.  Other apertures {were} then added to cover the full spatial extent of the [FeII] line, the most spatially extended emission line.  
The radii of these apertures are between 2 and 4 pixels depending on the signal to noise.  
Fig.~\ref{fig:lineratioaps} displays the final set of apertures overplotted on all 4 line emission maps.

In the Jn2 filter, the bandpass is so narrow that a global continuum fit is difficult to perform consistently for all aperture spectra.  Instead, a local continuum around each line {was} estimated by taking an average of the counts in wavelength channels on either side of the emission line.  This method {was} applied to both the Jn2 and Kbb data to maintain consistency between the bands.  The default channels used for the local continuum calculation are -450 to -300 km s$^{-1}$ and +300 to +450 km s$^{-1}$, the same velocity range used for the line S/N estimation described in Sec.~\ref{sec:linesnr}.  
The flux of each line {was} then determined by summing the continuum-subtracted counts in wavelength channels containing the emission line, by default $\pm$250 km s$^{-1}$ relative to the systemic velocity of the galaxy.  This velocity range was chosen to cover the wings of all 4 emission lines, including in the case that the lines are doppler shifted with respect to each other or with respect to the systemic velocity.  
In each aperture, the same continuum and line flux velocity ranges {were} used for all 4 emission lines so that the [FeII]/Pa$\beta$ and H$_2$/Br$\gamma$ line ratios are calculated using consistently-measured line fluxes and can therefore be compared to each other.  

If an emission line {was} not detected in an aperture, then an upper limit is placed on its line flux.  This upper limit {was} only needed for the hydrogen recombination lines, since these lines are much lower S/N than the [FeII] and 1-0 S(1) H$_2$ lines.  An upper limit on the Pa$\beta$ and Br$\gamma$ lines corresponds to a lower limit on the [FeII]/Pa$\beta$ and H$_2$/Br$\gamma$ line ratios.  
A conservative 3-$\sigma$ upper limit {was} calculated by taking the flux from a Gaussian emission line with a peak of 3 times the noise estimate described in Sec.~\ref{sec:linesnr} and a velocity dispersion of 100 km s$^{-1}$.  
This dispersion value {is} on the high end of the observed velocity dispersions of the [FeII] and H$_2$ emission lines (see Fig. \ref{fig:veldispmaps}), giving a conservative upper limit for the line flux. The flux of the Gaussian {was} computed by sampling the Gaussian at wavelength channels within the $\pm$250 km s$^{-1}$ velocity range and taking the sum of these values.  

Table \ref{tab:lineratio_results} gives the locations and radii of all 15 apertures, the velocity ranges used for the continuum and line flux estimation, and the resulting line ratios measurements or lower limits.  We find that [FeII] and 1-0 S(1) H$_2$ lines are detected in all 15 apertures and Pa$\beta$ and Br$\gamma$ are detected in 9 and 3 apertures respectively.  The lowest values of both line ratios ([FeII]/Pa$\beta$ = 2.5 and H$_2$/Br$\gamma$ = 1.2) are found in the aperture centered on the peak of the Pa$\beta$ and Br$\gamma$ emission.  Six of the other apertures with Pa$\beta$ detections are located to the north of the nucleus and show [FeII]/Pa$\beta$ ratios between 5.3 and 6.5.  Two other apertures to the south of the nucleus also have Pa$\beta$ detections with slightly lower values of 3.8 and 4.3, though these detections have lower S/N than the northern apertures.  
Two of the northern apertures also have detections of Br$\gamma$ and show H$_2$/Br$\gamma$ ratios of 2.4 and 2.7.

\begin{figure}
\begin{center}
\includegraphics[width=16cm]{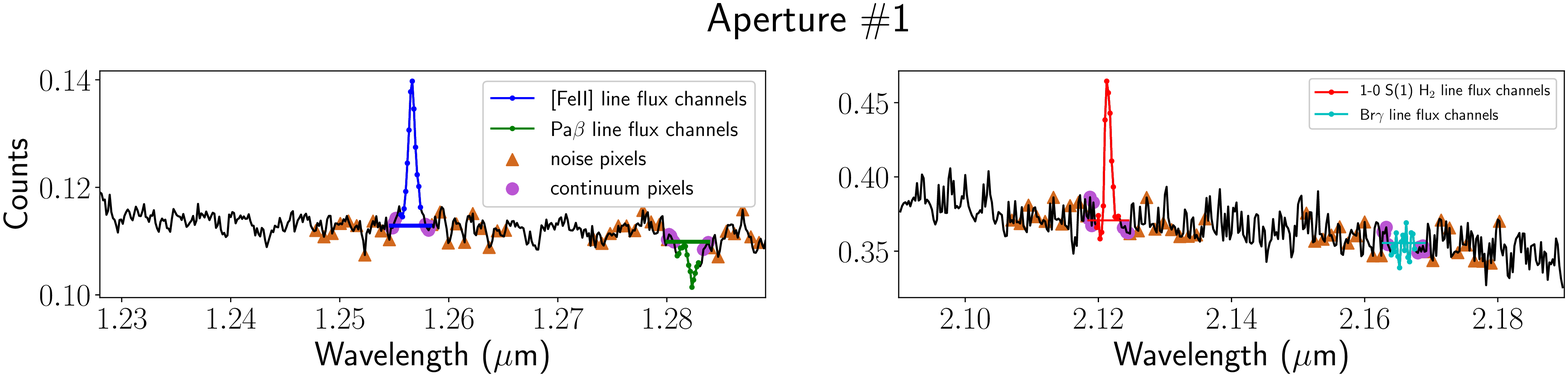}   
\includegraphics[width=16cm]{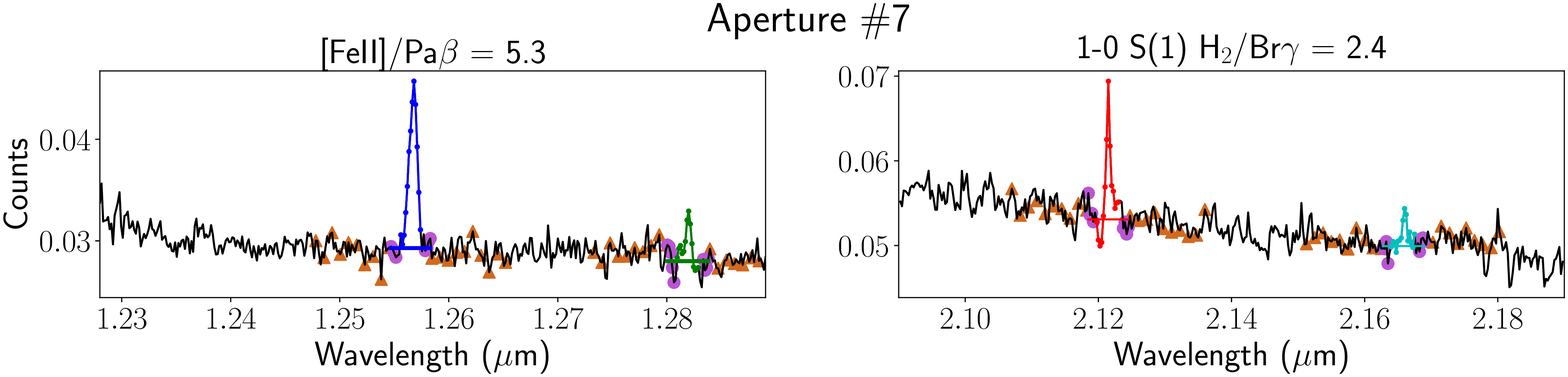}   
\includegraphics[width=16cm]{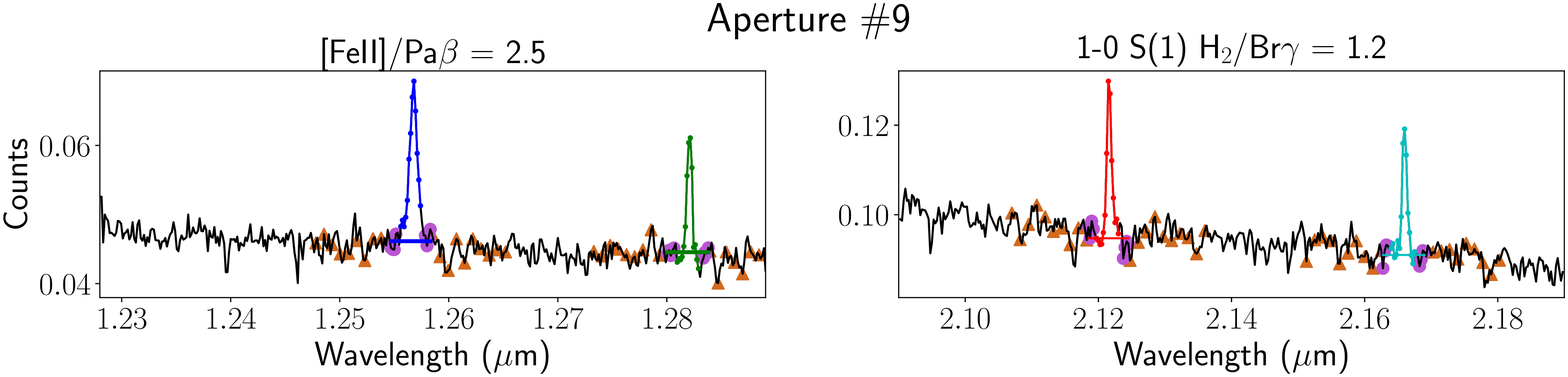}   
\includegraphics[width=16cm]{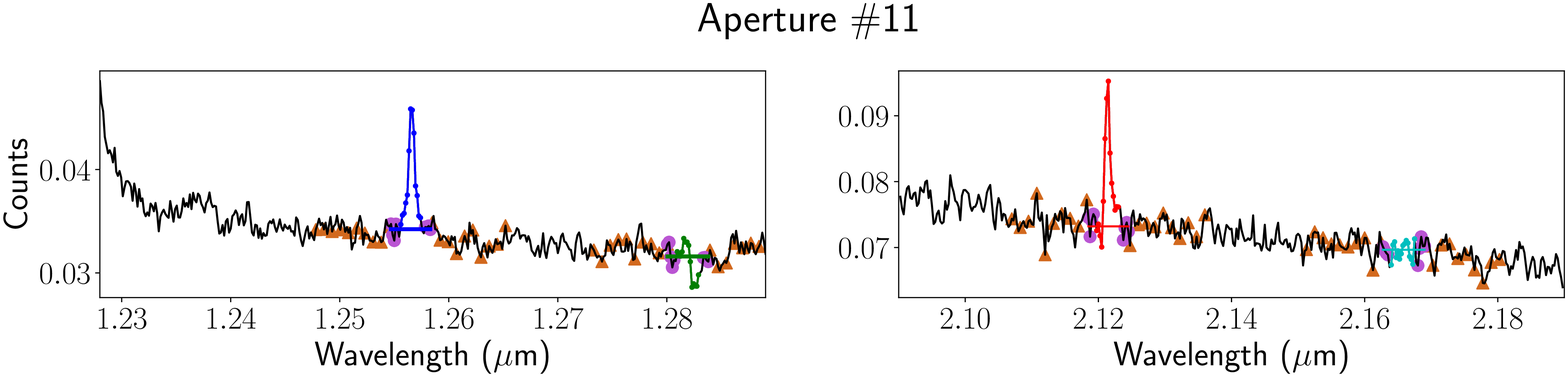}   
\includegraphics[width=16cm]{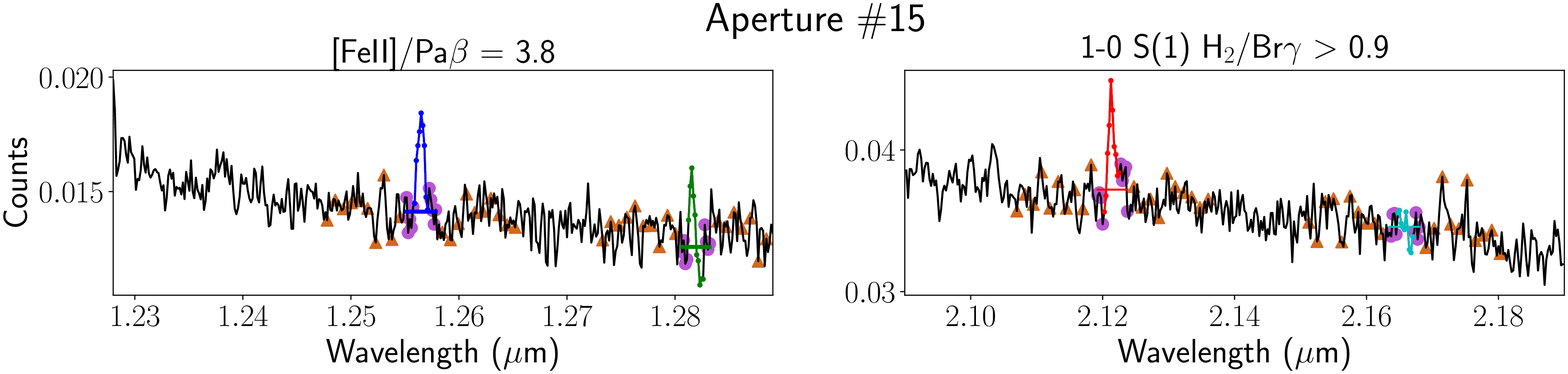} 
\end{center}
\caption
{ \label{fig:repspecs} Example spectra to represent the variety of emission line morphologies and ratios observed in NGC 404.  
The overall spectrum is plotted in black and each emission line is plotted in a different color to indicate the wavelength channels used to measure the line flux.  The pixels used for both the noise and the continuum estimation are also highlighted as brown triangles and purple circles respectively.  The continuum estimation is also plotted as a horizontal line spanning the channels used for both the continuum and line flux measurements.  
The spectra plotted here show examples of high (\#7) and low (\#9) [FeII]/Pa$\beta$ and H$_2$/Br$\gamma$ line ratios, a region in which Pa$\beta$ is detected but only a limit can be placed on Br$\gamma$ (\#15), and regions of strong (\#1) and weak (\#11) Pa$\beta$ absorption in which the line ratios are not constrained.
}
\end{figure}

Additionally, there are 6 apertures in which Pa$\beta$ absorption is detected and for these apertures no line ratio measurement or upper limit is reported. The Pa$\beta$ absorption is accompanied by either very weak to no emission (apertures \#1 and 12) or by strong emission (apertures \#2, 11, 13, and 14).  Fig.~\ref{fig:repspecs} displays 5 representative spectra to show the variation seen within the center of NGC 404.  The example spectra are:
the aperture centered on the nucleus of NGC 404 showing Pa$\beta$ absorption and no emission (\#1), 
a northern aperture with both Pa$\beta$ and Br$\gamma$ detections (aperture \#7), 
the aperture centered on the Pa$\beta$/Br$\gamma$ emission (\#9), 
an aperture with simultaneous detection Pa$\beta$ absorption and strong emission (\#11), and 
a southern aperture with Pa$\beta$ detected (\#15).

\begin{landscape}
\begin{deluxetable}{lrrrrrrrrrrrr}
\tabletypesize{\scriptsize}
\tablewidth{0pt}
\tablecaption{[FeII]/Pa$\beta$ and 1-0 S(1)/Br$\gamma$ Line Ratio Measurements in NGC 404}
\tablehead{
  \colhead{Aperture} & 
  \multicolumn{2}{c}{Aperture Location (arcsec)} & 
  \colhead{Radius} & 
  \multicolumn{2}{c}{Velocity Ranges (km sec$^{-1}$)\tablenotemark{a}} & 
  \multicolumn{4}{c}{Line Fluxes ($\times 10^{2}$ counts)} & 
  \colhead{[FeII]/} & 
  \colhead{1-0 S(1) H$_2$/} \\ 
  \colhead{Number} & 
  \colhead{($\Delta$ RA)} & 
  \colhead{($\Delta$ DEC)} & 
  \colhead{(pixels)} & 
  \colhead{Line Flux} & 
  \colhead{Continuum\tablenotemark{b}} & 
  \colhead{[FeII]} & 
  \colhead{Pa$\beta$} & 
  \colhead{1-0 S(1) H$_2$} & 
  \colhead{Br$\gamma$} & 
  \colhead{Pa$\beta$\tablenotemark{c}} & 
  \colhead{Br$\gamma$\tablenotemark{c}} &
}
\startdata
\#1 & $0.00$ & $0.00$ & 4.0 & $\pm250$ & 300-450 & $15$ & $-$ & $38$ & $-$ & $-$ & $-$ \\ 
\#2 & $-0.22$ & $-0.28$ & 4.0 & $\pm200$ & 200-350 & $5.0$ & $-$ & $9.9$ & $-$ & $-$ & $-$ \\ 
\#3 & $-0.10$ & $-0.53$ & 4.0 & $\pm175$ & 175-325 & $2.3$ & $0.5$ & $5.5$ & $<2.2$ & $4.3$ & $>2.6$ \\ 
\#4 & $0.00$ & $0.56$ & 4.0 & $\pm250$ & 300-450 & $4.4$ & $0.8$ & $3.6$ & $1.3$ & $5.7$ & $2.7$ \\ 
\#5 & $0.23$ & $0.49$ & 3.0 & $\pm250$ & 300-450 & $4.1$ & $0.6$ & $4.6$ & $<3.4$ & $6.4$ & $>1.4$ \\ 
\#6 & $-0.09$ & $0.35$ & 2.0 & $\pm250$ & 300-450 & $8.4$ & $1.4$ & $4.2$ & $<3.8$ & $6.0$ & $>1.1$ \\ 
\#7 & $0.09$ & $0.35$ & 2.5 & $\pm250$ & 300-450 & $9.6$ & $1.8$ & $4.3$ & $1.7$ & $5.3$ & $2.4$ \\ 
\#8 & $-0.16$ & $0.20$ & 3.0 & $\pm250$ & 300-450 & $10$ & $1.6$ & $13$ & $<4.5$ & $6.5$ & $>2.8$ \\ 
\#9 & $0.02$ & $0.21$ & 2.0 & $\pm250$ & 300-450 & $13$ & $5.4$ & $12$ & $11$ & $2.5$ & $1.2$ \\ 
\#10 & $0.20$ & $0.20$ & 3.0 & $\pm175$ & 175-325 & $7.2$ & $1.4$ & $4.0$ & $<5.1$ & $5.3$ & $>0.8$ \\ 
\#11 & $-0.26$ & $-0.02$ & 3.5 & $\pm250$ & 300-450 & $5.9$ & $-$ & $8.4$ & $-$ & $-$ & $-$ \\ 
\#12 & $0.26$ & $-0.04$ & 3.5 & $\pm250$ & 300-450 & $7.7$ & $-$ & $9.2$ & $-$ & $-$ & $-$ \\ 
\#13 & $0.06$ & $-0.28$ & 4.0 & $\pm250$ & 300-450 & $8.2$ & $-$ & $9.6$ & $-$ & $-$ & $-$ \\ 
\#14 & $0.29$ & $-0.34$ & 2.5 & $\pm200$ & 200-350 & $5.1$ & $-$ & $3.2$ & $-$ & $-$ & $-$ \\ 
\#15 & $0.18$ & $-0.55$ & 4.0 & $\pm175$ & 175-325 & $2.1$ & $0.5$ & $2.6$ & $<3.0$ & $3.8$ & $>0.9$
\enddata 
\label{tab:lineratio_results}
\tablenotetext{a}{All velocity values are relative to the systemic velocity of the galaxy (-48 km sec$^{-1}$).}
\tablenotetext{b}{The continuum velocity range given here refers to both the red (e.g., $+300$ to $+450$ km sec$^{-1}$) and blue (e.g., $-450$ to $-300$ km sec$^{-1}$) sides of the emission line.  Wavelength channels in these two ranges are used to estimate the continuum.}
\tablenotetext{c}{{Line} ratios measurements/upper limits are not reported for spectra with measured Pa$\beta$ absorption.}

\end{deluxetable}
\end{landscape}

\clearpage

\section{Discussion}
\label{sec:discussion}

The analysis and results in the previous section shows that the five dominant spectral features (infrared continuum, hydrogen recombination lines, molecular hydrogen rotation-vibrational lines, CO bandhead absorption, and the [FeII] emission line) all have distinct spatial and spectral structures and imply a complex nuclear environment within NGC 404. Indeed, the hydrogen recombination lines alone show both areas of emission and net absorption. In this section, we interpret the spectral properties of these five features in order to constrain the physical environment in the nucleus of NGC 404.  In particular, we discuss constraints on the nuclear stellar population (Sec.~\ref{sec:stellarpopdisc}) and the possible physical mechanisms responsible for exciting the H$_2$ (Sec.~\ref{sec:h2disc}) and the [FeII] gas emission (Sec.~\ref{sec:excitationmechanisms}).

\subsection{Stellar Population Constraints: CO and Pa$\beta$ Absorption and Pa$\beta$ Emission}
\label{sec:stellarpopdisc}

The stellar population in the nucleus of NGC 404 contributes both to the infrared continuum and to the CO and hydrogen recombination absorption features measured in the OSIRIS observations of NGC 404.  The infrared continuum observed at both J band and K band is centrally peaked, as shown in Fig.~\ref{fig:reducedcubes}.  This continuum emission likely arises from a nuclear stellar population that is centered at this peak of continuum emission, which also coincides with the peak of the CO bandhead and Pa$\beta$/Br$\gamma$ absorption.  
A nuclear star cluster was identified in NGC 404 by \citet{2001AJ....122..653R} and its stellar population has been been studied at optical wavelengths \citep{2010Seth, 2017ApJ...836..237N}. 
 Most recently, \citet{2017ApJ...836..237N} used the velocity field derived from the CO bandhead feature observed with the Gemini/NIFS IFS to derive an upper limit on the mass of the putative central BH.  While deriving a BH mass from the data presented here is outside the scope of this work, we do find a stellar velocity field consistent with these prior studies when considering the smaller field of view of the OSIRIS data.

The CO bandhead stellar absorption feature present in the Kbb data cube can be used to constrain the dominant age of the central stellar population in NGC 404.  This absorption feature is sensitive to intermediate age stellar populations because the CO bandheads are stronger in cooler stars with lower surface gravity (i.e., giant and supergiant stars) than in dwarf stars.  
The strength of the absorption feature is often characterized by the CO index.  We {employed} the CO index defined by \citet{2008A&A...489..885M}, $D_{CO}$.  That work found the $D_{CO}$ index to be less sensitive to velocity dispersion, radial velocity, and low S/N spectra than previously-defined CO indices.  $D_{CO}$ is computed by taking a ratio of the average flux in two continuum bands (2.2460 -- 2.2550 $\mu$m and 2.2710 -- 2.2770 $\mu$m) over the average flux in one absorption band (2.2880 -- 2.3010 $\mu$m), such that a $D_{CO}$ value of $\sim$1 indicates no CO absorption and increasing values indicate a deeper CO absorption depth relative to the continuum.  For reference, individual dwarf stars have $D_{CO}$ values ranging from $\sim$1 -- 1.1, red supergiants have values of $\sim$1.2 -- 1.25, and AGB stars have values up to $\sim$1.3.

\begin{figure}[t]
\begin{center}
 \begin{tabular}{cc}
 \includegraphics[height=7cm]{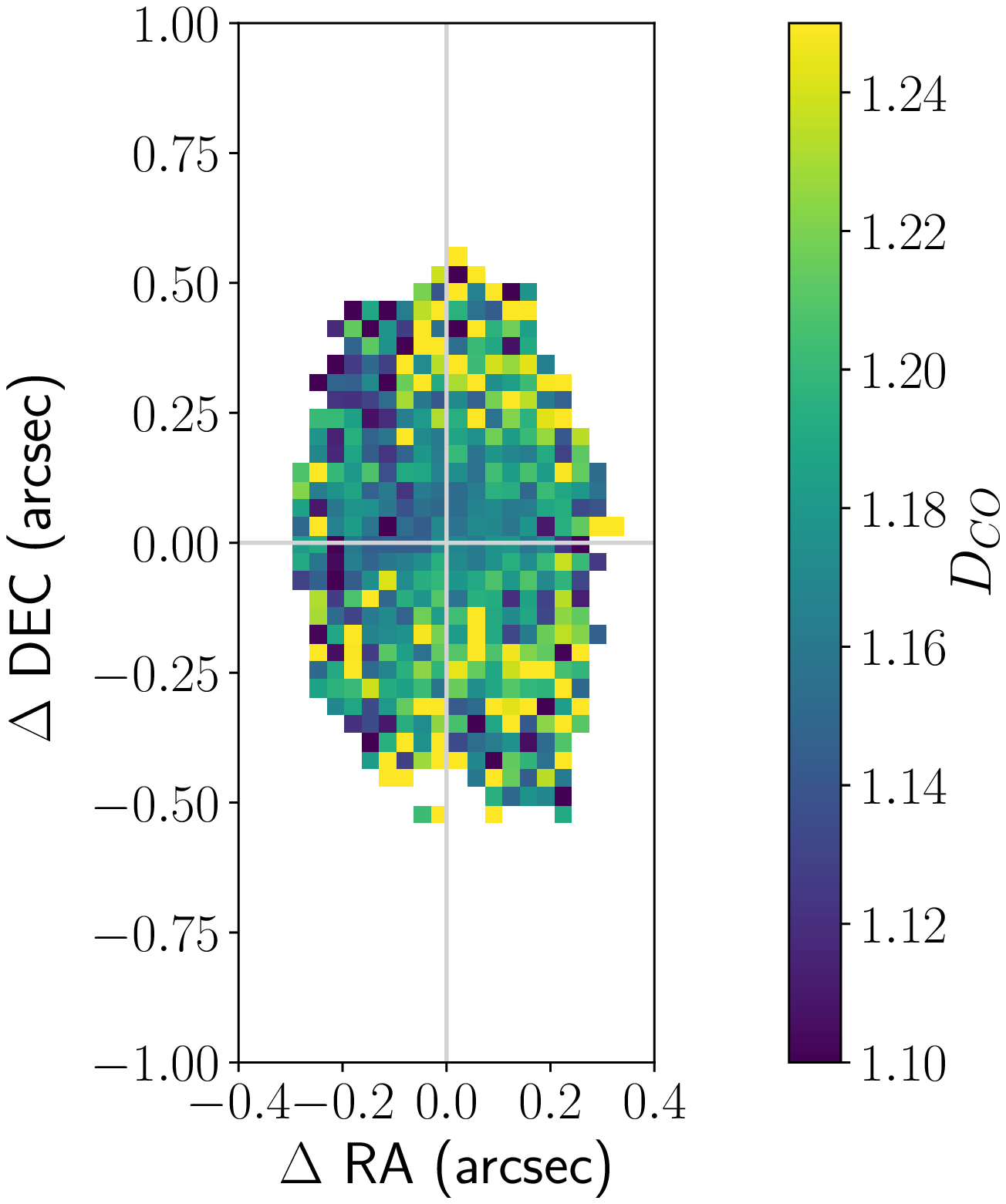} &
 \includegraphics[height=7cm]{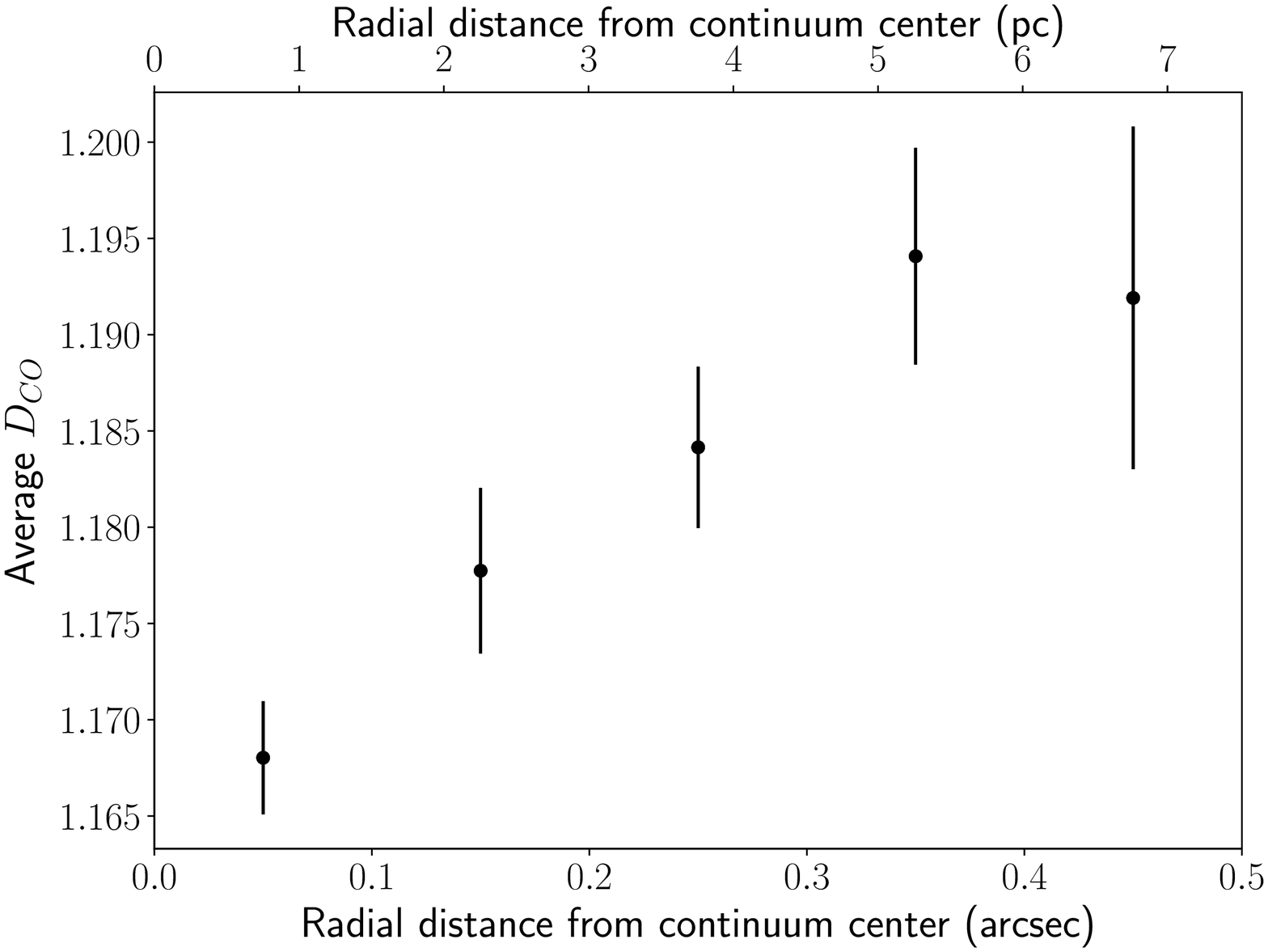} \\
 \end{tabular}
 \end{center}
 \caption
{ \label{fig:COindexmap} Map of the CO index $D_{CO}$ across the central $\sim$0.5 by 1.0 arcsec (7.5 by 15 pc) of the nucleus of NGC 404 (left) and $D_{CO}$ versus radius (right).  In the spatial map, only the spaxels with a continuum S/N greater than 7.0 are displayed.  The origin of the map is set to the center of the Kbb continuum emission and is highlighted by the horizontal and vertical gray lines.  The $D_{CO}$ index in the nucleus of NGC 404 has a value of 1.168 $\pm$ 0.003 at the center and increases to values of $\sim$1.195, or $\sim$25\% more CO absorption, 
at distances of $\pm$5-7 pc from the center.}
\end{figure}

The $D_{CO}$ index in the nucleus of NGC 404 {was} computed for each individual spaxel of the Kbb data cube by first shifting the wavelength ranges of the continuum and absorption bands to the systemic velocity of NGC 404.  The radial dependence of the CO index {was} also probed by taking the average of the CO index in 0.1 arcsec (1.5 pc) wide radial bins from 0 to 0.5 arcsec (7.5 pc).  The error in each radial bin {was} taken as the error of the mean (standard deviation divided by a square root of the number of pixels contributing to the average). 
The resulting map and radial dependence of the $D_{CO}$ index is shown in Fig.~\ref{fig:COindexmap} for spaxels with a continuum S/N greater than 7.0.  
We measure a significant spatial variation of the CO index, with a value of 1.168 $\pm$ 0.003 at the center of NGC 404 and increasing to values of $\sim$1.195 ($\sim$25\% more CO absorption) 
at 5-7 pc from the Kbb continuum center.

The strength of the CO absorption quantified by the $D_{CO}$ index can be interpreted by comparing indices measured on the spectra of a simulated stellar population with the indices measured from the data, thereby constraining the age of the stellar population. \citet{2015A&A...582A..97M} performed single age and metallicity stellar population synthesis modeling in the NIR based on empirical stellar spectra from the IRTF spectral library \citep{2009ApJS..185..289R}.  That work {computed} a number of spectral indices and {tracked} their values as a function of age (from 1 - 14 Gyr) and metallicity (-0.70 to +0.20 dex relative to solar) of the stellar population.  Note that earlier ages {were} not modeled due to the lack of high temperature main sequence stars in the IRTF spectral library.  

When comparing the measured $D_{CO}$ values to the simulated values \citep[Fig.~11 in][]{2015A&A...582A..97M}, we find that the $D_{CO}$ value at the nucleus of 1.168 $\pm$ 0.003 is lower than most of the computed models and can only be produced by a stellar population with an age of 1 - 1.5 Gyr and a metallicity of 0 to +0.2 dex relative to solar.  The higher value of $\sim$1.195 found at larger radii from the nucleus is reproduced by a larger set of models with ages ranging from 1 to 7 Gyr and with metallicities spanning the full sampled range.  
The dominant stellar population age of 1 -- 1.5 Gyr that we find in the central 1.5 pc of NGC 404 is consistent with previous results using optical spectroscopy \citep[e.g.,][]{2010Seth, 2017ApJ...836..237N}.

We also consider the constraints that the Pa$\beta$ absorption feature observed in the central $\sim$4 pc of NGC 404 places on the stellar population.  This absorption feature can be affected by contamination of emission unlike the CO bandheads, but it is still interesting to explore the possible stellar population ages consistent with the measurement of this feature's equivalent width (EW).  
{Since stellar populations models such as \citet{2015A&A...582A..97M} do not give predictions for Pa$\beta$ EW versus age, we compared} the measured EW in a spatial aperture centered on the nucleus of NGC 404 with a 2 pc radius (aperture \#1 in Fig.~\ref{fig:h2_aperture_spectra} and \ref{fig:lineratioaps}) to those measured from the telluric standard spectra of A0V and G2V stars (see Sec.~\ref{sec:observations}) and to other dwarf stars ranging from spectral type F to M stars from the IRTF spectral library \citep{2009ApJS..185..289R}.  
{We find an EW consistent with a stellar population dominated in the near-infrared by late F/early G stars or by late B stars.  
Based on the main sequence lifetimes of these stars, this Pa$\beta$ EW implies the presence of stars with an age of $<5 - 8$ Gyr, consistent with our CO bandhead constraints, or a young stellar population with an age of $<500$ Myr. 
We also compared the measured EW to that of a nominal old stellar population dominated by K0 stars to determine what fraction of A stars is needed to reproduce the measured Pa-$\beta$ absorption. Using the Pa$\beta$ EW of K0 and A0 stars and adopting the relative J band magnitudes of these stars from \citet{2013ApJS..208....9P}\footnote{See also: \url{http://www.pas.rochester.edu/~emamajek/EEM_dwarf_UBVIJHK_colors_Teff.txt}}, we find that an A star number fraction of $\sim$1\% (equivalent to a mass fraciton of $\sim$4\%) is required to reproduce the measured EW.  This result suggests that there is a young component in the nuclear stellar population of NGC 404 in addition to the $\sim$1 Gyr component indicated by the CO bandhead absorption.
The previous optical spectroscopic studies mentioned above have also found a component of young stars in the nucleus of NGC 404, with the preferred stellar population model in the most recent work finding that this young component (<500 Myr) is a small fraction of the total mass ($\sim$6\%) compared to the 1 Gyr stars \citep[][]{2017ApJ...836..237N}.  
}
Those authors also pointed out a region of blue optical colors slightly offset to the west of the nucleus that they attribute to this young stellar population, which coincides with the Pa$\beta$ absorption seen in the velocity-resolved emission maps offset $\sim$1 pc to the west of the nucleus (see 50 and 100 km s$^{-1}$ velocity channels in Fig.~\ref{fig:velchannelmaps}).  Another sign of the presence of younger stars in this region is the Pa$\beta$/Br$\gamma$ emission seen 3 pc to the north of the nucleus of NGC 404.  This compact clump of hydrogen recombination emission is spatially coincident with bright, extended [FeII] emission (see 50 km s$^{-1}$ channel in Fig.~\ref{fig:velchannelmaps}) and likely originates from an HII region powered by star formation.

\subsection{Molecular Hydrogen Gas Disk}
\label{sec:h2disc}

The nuclear molecular gas in NGC 404 is observed through the H$_2$ ro-vibration emission lines.  In this emission line, we observe a central peak of emission in the central $\sim$5 pc that is coincident with the peak of stellar infrared continuum, as well clumps of emission to the southeast of the continuum peak {that are either physically separate knots of molecular gas or hot spots in a larger rotating disk}.  The H$_2$ velocity profile is distinct from the stellar velocities (Fig.~\ref{fig:CO_gas_velcomparison}) and shows signs of rotation at a position angle of -45$^{\circ}$ spanning line-of-sight velocities of $\pm$30 km s$^{-1}$ across both the central and offset clumps of H$_2$ line emission (Fig.~\ref{fig:veldispmaps}).  Similar to the stellar velocities derived from the CO bandhead, the molecular gas velocities have previously been used to constrain the mass of the possible central BH \citep{2010Seth}, although a more recent work found that the kinematics are dominated by non-gravitational motions and thus do not give a reliable BH mass measurement \citep{2017ApJ...836..237N}.  We find {H$_2$} gas velocities in the OSIRIS data that are consistent with the {H$_2$ velocity maps derived from the Gemini/NIFS data} presented in those previous works.

The dominant excitation mechanism of the H$_2$ molecular gas can be constrained by comparing H$_2$ line ratios derived in Sec. \ref{sec:H2temps} to models.    
These line ratios can also be converted to two excitation temperatures: the rotation temperature ($T_{rot}$, derived from the 1-0 S(2)/1-0 S(0) ratio) and the vibrational temperature ($T_{vib}$, derived from the 1-0 S(1)/2-1 S(1) ratio). 
We {applied} the equations from \citet{2002MNRAS.331..154R}, which assume local thermal equilibrium and therefore a thermal distribution for the ro-vibrational level populations:
\begin{equation}
T_{rot} = \frac{-1113}{\ln \left( 0.323 \times \frac{\textrm{1-0 S(2)}}{\textrm{1-0 S(0)}}\right)}
\end{equation}
 \begin{equation}
 T_{vib} = \frac{5600}{\ln \left( 1.355 \times \frac{\textrm{1-0 S(1)}}{\textrm{2-1 S(1)}}\right)}
\end{equation}
In the case of purely thermal (collisional) excitation, the molecular gas is in local thermal equilibrium and the level populations are set by a Boltzmann distribution.  The two excitation temperatures are therefore equal to each other and to the kinetic temperature of the gas.  
For UV fluorescence excitation, upper energy levels of the H$_2$ are over populated compared to the Boltzmann distribution due to the absorption of UV photons, resulting in a higher $T_{vib}$ ratio compared to $T_{rot}$ by a factor of $\sim$6 \citep{1987ApJ...322..412B}.

\begin{figure}[t]
\begin{center}
  \includegraphics[width=12cm]{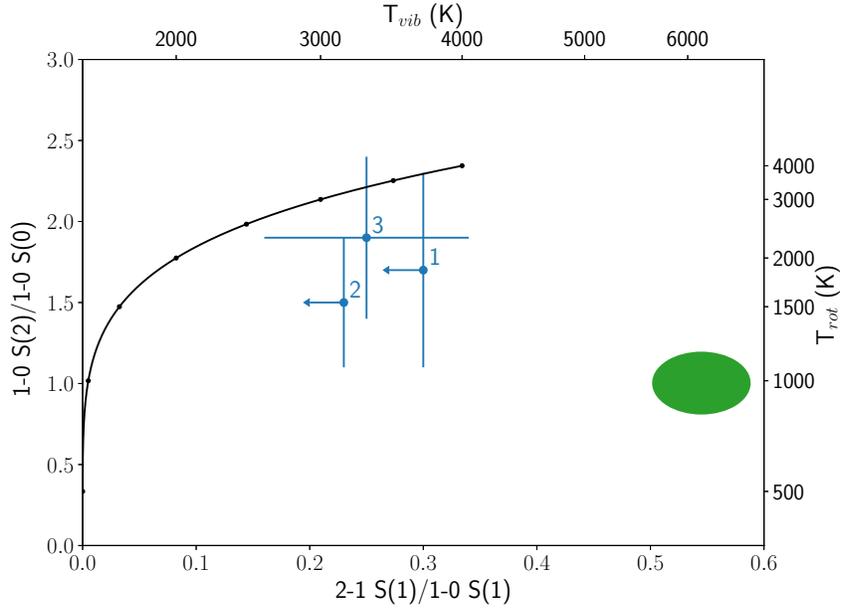}
\end{center}
\caption
{ \label{fig:h2excitation} 2-1 S(1)/1-0 S(1) versus 1-0 S(2)/1-0 S(0) H$_2$ line ratios for each of the 3 spatial apertures displayed in Fig.~\ref{fig:h2_aperture_spectra} \citep{1994ApJ...427..777M}.  The excitation temperatures $T_{rot}$ and $T_{vib}$ calculated from each line ratio are also shown on the top and right axes.  The black curve shows the line ratios and temperatures corresponding to thermal excitation ($T_{rot} = T_{vib}$) for gas temperatures ranging 500 to 4000 K, with black points having a spacing of 500 K.  The green ellipse covers the range of H$_2$ spectral models from UV fluorescence excitation from \citet{1987ApJ...322..412B}.  The line ratios in the three apertures are all consistent with each other and fall near the thermal excitation curve at temperatures of 2000 - 4000 K with a slight shift towards the UV fluorescence models.  These results indicate that the H$_2$ gas in NGC 404 is thermally excited with some contribution from UV fluorescence.    
}
\end{figure}

We {compared} the measured line ratios of 2-1 S(1)/1-0 S(1) and 1-0 S(2)/1-0 S(0) and the corresponding excitation temperatures of $T_{vib}$ and $T_{rot}$ in each of the three spatial apertures in NGC 404 in Fig.~\ref{fig:h2excitation} \citep{1994ApJ...427..777M}.  These measurements {were} compared to expected values for thermal excitation ($T_{vib}$ = $T_{rot}$) and UV fluorescence excitation \citep[$T_{vib}$ much greater than $T_{rot}$;][]{1987ApJ...322..412B}.  We {measured} consistent line ratios and excitation temperatures within the three spatial apertures 
centered on the central bright H$_2$ emission and the two peaks of emission offset to the southeast of the nucleus.   
In the  2-1 S(1)/1-0 S(1) versus 1-0 S(2)/1-0 S(0) parameter space in Fig.~\ref{fig:h2excitation}, the line ratio measurements and/or upper limits in each aperture fall near the thermal excitation curve of $T_{vib}$ = $T_{rot}$ with an offset towards the UV fluorescence excitation models.  
We therefore conclude that the H$_2$ excitation mechanism is consistent with thermal excitation at all three spatial locations, with a temperature of 2000-4000 K, plus some contribution of UV fluorescent excitation.  Given the kinematic and morphological structure of the H$_2$ emission, we interpret the central peak of emission as a rotating disk of thermally excited molecular gas.  
Thermally excited disks of H$_2$ gas have been commonly found at the centers of classical Seyferts \citep[e.g.,][]{2009MNRAS.394.1148S, 2009ApJ...696..448H, 2013ApJ...768..107H, 2015MNRAS.451.3587R, 2016IAUS..319...59H} as well as optically-classified LINERs and low luminosity AGN \citep[e.g.,][]{2013ApJ...763L...1M, 2013MNRAS.428.2389M, 2014MNRAS.438.2036M}.  
{The peaks of offset H$_2$ emission are also dominated by thermal excitation, but it is unclear whether these peaks trace hot spots in a larger scale rotating disk associated with the central emission peak or are physically separate clumps of gas. }

Although we can determine that the H$_2$ gas in NGC 404 is dominated by thermal excitation, it is more difficult to constrain the physical mechanism heating the H$_2$ gas from the line ratios alone.  
One possibility is that the H$_2$ is being heated by the same shock front that is exciting the [FeII] emission as these shocks impinge on the central disk of molecular gas (discussed in detail in Sec.~\ref{sec:excitationmechanisms}).  Other potential heating sources include UV radiation from young stars or X-ray radiation from X-ray binaries or the central weakly accreting BH. 
To explore what might be heating the gas, we {constrained} the mass of the warm H$_2$ gas and its total internal energy.  We {used} the equation presented in \citet{2013MNRAS.428.2389M} to convert the 1-0 S(1) line flux to the warm H$_2$ gas mass ($M_{H_2}$), which assumes a kinetic gas temperature of 2000 K:
\begin{equation}
M_{H_2} \simeq 5.0875~\times~10^{13}~\left(\frac{D}{\textrm{Mpc}}\right)^2   \left(\frac{\textrm{1-0 S(1)}}{\textrm{erg s$^{-1}$ cm$^{-2}$}}\right) 10^{0.4~\times~A_{2.2}}.
\end{equation}
In this equation, D is the distance to the galaxy (3.1 Mpc for NGC 404) and $A_{2.2}$ is the extinction at 2.2 $\mu$m.  We {used} the 1-0 S(1) H$_2$ line flux presented in \citet{1998Larkin} for the central $\sim$10 by 45 pc of NGC 404 and {assumed} a large value of 1.0 for $A_{2.2}$ (corresponding to a visible extinction $A_{V}$ of $\sim$10.0) to find a conservative upper limit for the H$_2$ gas mass.  
The resulting maximum warm molecular gas mass in the nucleus of NGC 404 is 3 M$_\odot$.
This warm H$_2$ mass is comparable to masses derived by \citet{2013MNRAS.428.2389M} for a sample of 6 nearby galaxies that exhibit a range of nuclear activity, when taking into account the different spatial scales probed by those data compared to \citet{1998Larkin}. 
Note that this is only a measurement {of} the mass of the gas that is warm and emitting the thermally exciting lines; it does not include the cold H$_2$ gas, which is expected to be a factor of 10$^5$ to 10$^7$ higher than the warm gas mass \citep[see][]{2005ApJ...633..857D, 2006A&A...454..481M}.  
The warm H$_2$ mass can then be converted to the total internal energy using $U = c_{P}nT$, where c$_{P}$ is the specific heat at constant pressure of H$_2$, $n$ is the number of moles of H$_2$, and $T$ is the gas kinetic temperature of 2000 K.  Given a value of c$_{P}$ at 2000 K of 34 J / (K mol) from NIST Standard Reference Database, we find an upper limit for the total internal energy of the H$_2$ gas of $2\times10^{45}$ erg.  This total energy is 6 orders of magnitude lower than the total kinetic energy released by a single supernova.  
This low value of the total warm H$_2$ energy suggests that there are many possible physical mechanisms, including the supernova remnant shock, that could contribute to the heating of the gas and thereby cause the thermal excitation that we observe.  

\subsection{Shock Front Traced by [FeII] Gas Emission}
\label{sec:excitationmechanisms}

The [FeII] line emission has a distinct morphology and velocity profile from the stellar continuum and the molecular gas in the nucleus of NGC 404.  The emission from this ionized gas is spatially extended across the central $\sim$15 pc of the nucleus and follows a complex velocity structure that spans a wide range of velocities down to -150 km s$^{-1}$ and up to +200 km s$^{-1}$ (Fig.~\ref{fig:velchannelmaps}).  A key constraint on the important physical mechanisms in the nucleus of NGC 404 is through this shock-sensitive emission line, in particular the ratio of the [FeII] line flux over Pa$\beta$.  We argue below that the [FeII] emission is excited by shocks and {might be} tracing a shock front caused by a supernova.

\begin{figure}[th]
\begin{center}
 \includegraphics[width=13cm]{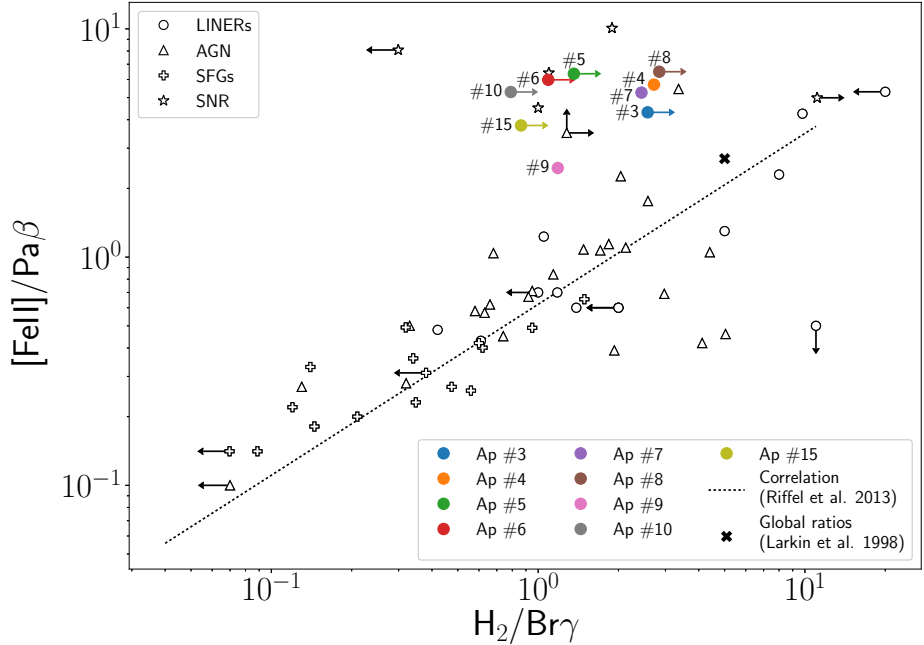} 
\end{center}
\caption
{ \label{fig:lineratiocorr} Plot of the [FeII]/Pa$\beta$ and 1-0 S(1) H$_2$/Br$\gamma$ emission line ratio measurements and lower limits for different spatial apertures.  Apertures in which Pa$\beta$ absorption is detected are not plotted.  The correlation between these two line ratios as measured from spatially-integrated slit spectra of AGN, star forming galaxies (SFGs), and LINER galaxies is shown as a black dashed line \citep{2013Riffel}.  Values from individual sources in the literature are plotted for LINERs, SFGs, AGN, and supernova remnants (SNRs) \citep{1998Larkin, 2004ApJ...601..813D, 2004A&A...425..457R, 2005MNRAS.364.1041R, 2013Riffel}.   The global line ratios for NGC 404 from \citet{1998Larkin} are plotted as a black cross.  We find that the line ratios in NGC 404 measured by OSIRIS are generally 1.5 - 2.5 times higher than the global line ratios measured by \citet{1998Larkin} and fall near the shocked regions of supernova remnants in this parameter space.  The aperture centered on the Pa$\beta$/Br$\gamma$ emission (\#9) is lower than the other apertures and is consistent with the global line ratios.  The high line ratios measured by OSIRIS are evidence that the [FeII] emission in NGC 404 is excited by widespread shocks in the nucleus of this galaxy.
}
\end{figure}

Constraints on the physical mechanism that excites the [FeII] emission in NGC 404 can be made using measurements of the [FeII]/Pa$\beta$ ratio (Sec.~\ref{sec:NIRdiagnosticratios}).  High values of this ratio ($>$2.0) are indicative of shock excitation, which can efficiently {destroy/ablate} dust grains and thereby increase the gas phase abundance of iron, which is typically highly depleted in the interstellar medium \citep{1994Blietz, 1998Larkin}.  
Fig.~\ref{fig:lineratiocorr} plots the [FeII]/Pa$\beta$ and H$_2$/Br$\gamma$ line ratios presented in Table \ref{tab:lineratio_results}.  These measurements, which probe physical scales of 1-2 pc, are compared to values in other objects from the literature and to the overall correlation found in these line ratios from spatially-integrated spectra of starburst, AGN, and LINER galaxies \citep{1998Larkin, 2013Riffel}.  
The global line ratios for NGC 404 derived by \citet{1998Larkin} from slit spectra that cover the central $\sim$10 by 45 pc are also plotted.  
We find that the [FeII]/Pa$\beta$ line ratio measured in the small apertures with OSIRIS is generally 1.5 - 2.5 times higher than the global [FeII]/Pa$\beta$ ratio measured by \citet{1998Larkin}.  The aperture centered on the Pa$\beta$/Br$\gamma$ emission feature (\#9) has a [FeII]/Pa$\beta$ line ratio that is lower than the rest of the spatial locations and consistent with the global ratio.  
The high [FeII]/Pa$\beta$ values ranging from 4 - 6.5 are among the highest measured in extragalactic sources and are similar to those seen in the shocked regions of supernova remnants, also plotted on Fig.~\ref{fig:lineratiocorr}.  From these high [FeII]/Pa$\beta$ ratios we conclude that the [FeII] emission in NGC 404 is due to shock excitation.

\begin{figure}[t]
\begin{center}
 \begin{tabular}{c}
 \includegraphics[width=8cm]{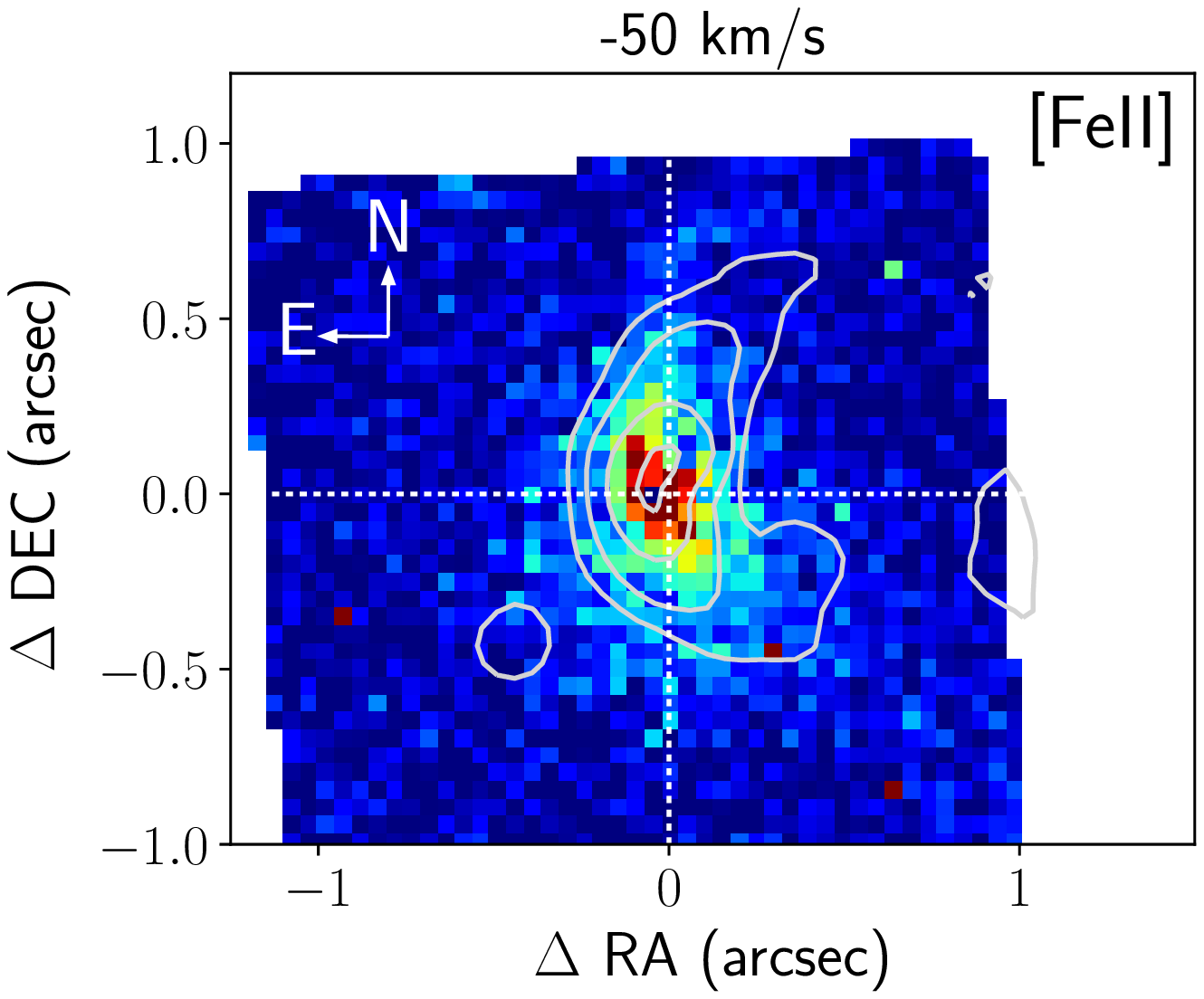} \\  
 \includegraphics[width=8cm]{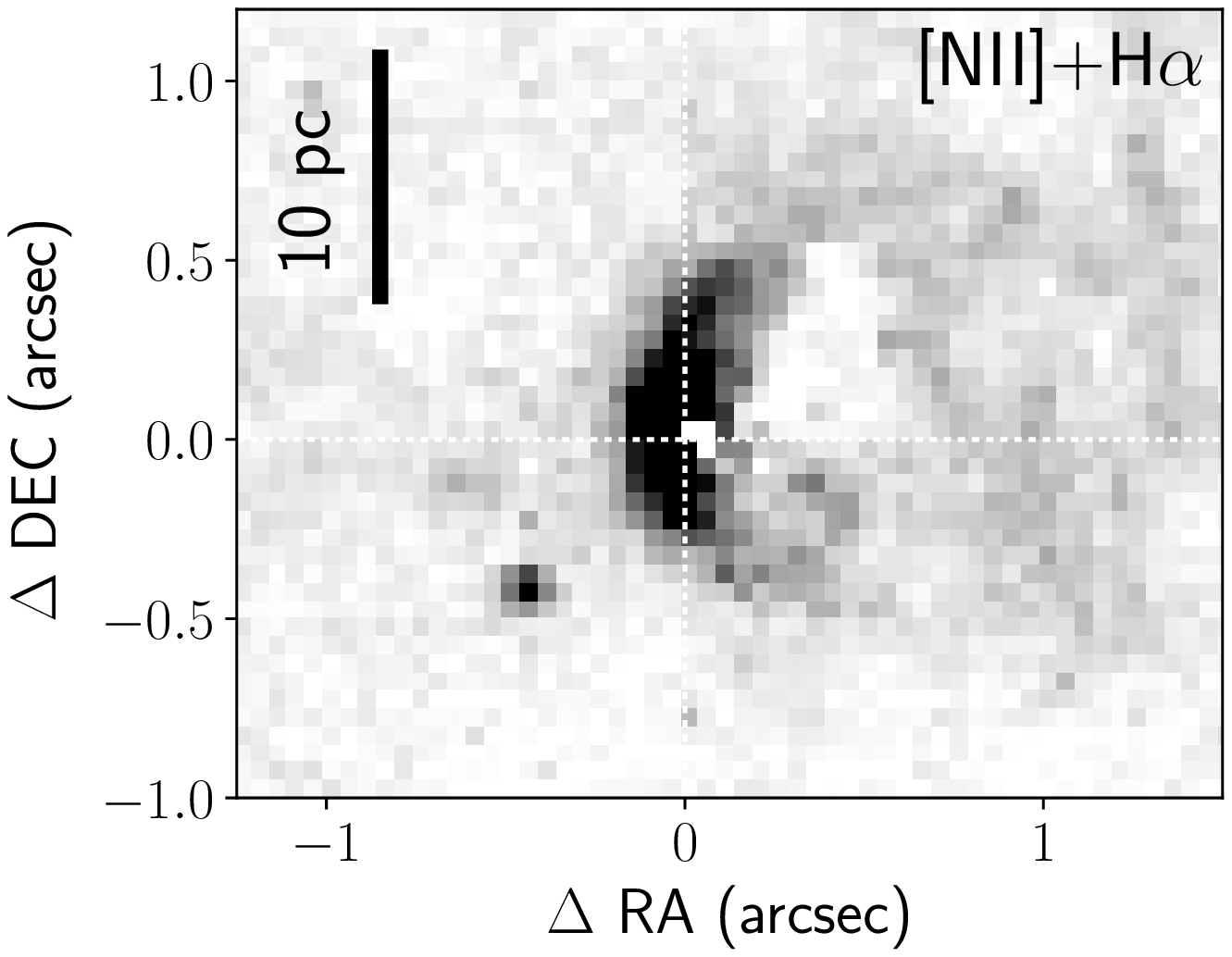}
 \end{tabular}
\end{center}
\caption
{ \label{fig:feII_hst_compare} Comparison of the [FeII] emission morphology observed at $-50$ km s$^{-1}$ relative to the systemic velocity of the galaxy (this work; see Sec.~\ref{sec:velchannelmaps}) and the [NII]+H$\alpha$ emission observed using HST narrowband imaging \citep{2000ApJ...532..323P}.  The origin of each plot is set to the continuum center as determined by a Gaussian fit to the spectrally-averaged Jn2 data cube (top) and to the F547M continuum filter image (bottom).  The [FeII] emission map {was} derived from the OSIRIS data taken in the 50 mas plate scale mode (see Sec.~\ref{sec:observations} for details).  Contours of the [NII]+H$\alpha$ image are overplotted on the [FeII] map, highlighting that the curvature and spatial extension of the [FeII] emission at this velocity is similar to the [NII]+H$\alpha$ structure.  Based on these similarities and the very high [FeII]/Pa$\beta$ ratios observed in NGC 404, {we argue that the [FeII] emission may be excited by shocks originating from a supernova that occurred at the center of the [NII]+H$\alpha$ bubble $\sim$5 pc to the west and are now impinging on the central dense gas in the nucleus of NGC 404.}  
}
\end{figure}

One possible source of the shocks that excite the [FeII] emission is one or more supernova remnants. \citet{1998Larkin} previously suggested this scenario for NGC 404's [FeII] emission and found that the [FeII] emission seen in the central $\sim$10 by 45 pc is only a few times that produced by a single supernova remnant.  
\citet{2000ApJ...532..323P} also noted that the extended, bubble-like structure of [NII]+H$\alpha$ emission seen in Hubble Space Telescope (HST) optical narrowband imaging is similar to that of a supernova remnant.  
In comparing the [NII]+H$\alpha$ imaging data to the [FeII] emission at $-50$ km s$^{-1}$, we find that the curvatures and spatial extension of the [FeII] emission to the north and southwest of the nucleus is very similar to the [NII]+H$\alpha$ structure (see Fig.~\ref{fig:feII_hst_compare}).  
The similarity between these morphologies and the high [FeII]/Pa$\beta$ ratios suggest that the [FeII] emission originates from a shock front caused by one or more supernovae which were offset to the west of the nucleus by $\sim$5 pc (i.e., the location of the center of the [NII]+H$\alpha$ bubble seen in the HST data).  
As this shock front reaches the denser gas in the nucleus, it can then excite the [FeII] emission and produce the observed [FeII]/Pa$\beta$ ratios. 

A supernova remnant was previously considered by \citet{2012ApJ...753..103N} as a possible source of the observed radio and hard X-ray \citep[2 - 10 keV;][]{2011ApJ...737...77B} nuclear emission in NGC 404.  That work explored a core-collapse supernova as the original source of the remnant and argued that this scenario was statistically unlikely given the current star formation rate. 
It is also possible, however, that the supernova causing the observed [FeII]-exciting shocks was instead a type Ia, which would be more consistent with the dominant stellar population age of $\sim$1 Gyr found in this work (see Sec.~\ref{sec:stellarpopdisc}) and in previous optical studies of the nucleus of NGC 404 at seeing-limited \citep{0004-637X-605-1-105, 2010A&A...513A..54B} and diffraction-limited angular resolutions \citep{2010Seth, 2017ApJ...836..237N}.  {For the nuclear star cluster mass from \citet{2017ApJ...836..237N} of 1.2 x 10$^7$ solar masses and the dominate age of $\sim$1 Gyr, the expected current rate of type Ia supernovae is $\sim$5 every 1 Myr \citep{2012PASA...29..447M}.  
Given this current rate of type Ia supernovae and assuming a lifetime of a supernova remnant on order 10,000 to 100,000 years, we used the Poisson distribution to calculate the probability that a supernova occurred in the last 10,000 to 100,000 years to be 5 - 30\%.  
It is therefore possible that we are currently observing a type Ia supernova remnant in the nucleus of NGC 404, and we argue that the dominant source of the shocks exciting the [FeII] emission could be due to such a supernova.
}

\section{Summary}

We have presented the first spatially-resolved data of J-band [FeII] and Pa$\beta$ emission in the nucleus of NGC 404 and have compared it to the K-band H$_2$ and Br$\gamma$ emission and CO stellar absorption.  These high angular resolution data reveal differences in morphology and kinematics of the stars and the ionized and molecular gas at pc spatial scales in this LINER nucleus.  
We find that the data are best explained by at least four physical components: a nuclear star cluster, an HII region, a thermally-excited rotating molecular gas disk, and a shock front traced by the [FeII] emission. We summarize our results related to these physical components below:
\begin{itemize}
	\item{Stellar population and HII region: We {observed} infrared continuum and the CO, Pa$\beta$, and Br$\gamma$ absorption features from the nuclear stellar population.  We {used} the depth of the CO absorption feature, quantified by the CO index, to constrain the stellar population and find a dominate age of $\sim$1 Gyr at the center of NGC 404, with a wider range of possible ages from 1 - 7 Gyr at increasing radii.  We also {identified} a region of Pa$\beta$ and Br$\gamma$ emission to 3 pc north of the nucleus that is likely an HII region powered by star formation.}
    \item{H$_2$ gas disk: The H$_2$ gas emission primarily originates from a central rotating disk whose emission spans the central $\sim$5 pc of the nucleus.  This disk is rotating at a position angle of -45$^\circ$ on the sky with line-of-sight rotation speeds up to $\pm$30 km s$^{-1}$.  We also find extended H$_2$ emission to the southeast of the nucleus that may originate from physically separate knots of gas or from hot spots in the rotating disk.  Ratios of the observed ro-vibration emission lines indicate that all the observed molecular gas emission is thermally excited with some contribution from UV fluorescence.  The exact source of the heating is not well constrained, but we find that a single supernova has 6 orders of magnitude more energy than required to have heated the gas to its current temperature.}
    \item{[FeII]-traced shock front: The observed [FeII] emission extends across the central $\sim$15 pc of the nucleus and has a complex velocity structure spanning out to -150 and 200 km s$^{-1}$.  High [FeII]/Pa$\beta$ line ratios across the spatial extent of this emission provide strong evidence that the [FeII] emission is excited by shocks.  {We argue that a possible physical source of the [FeII] shock excitation is a supernova remnant.}  This scenario is supported by both the high [FeII]/Pa$\beta$ ratios similar to the shocked regions of supernova remnants (Fig.~\ref{fig:lineratiocorr}) and the similar morphology of [FeII] and [NII]+H$\alpha$ from HST narrowband imaging \citep[Fig.~\ref{fig:feII_hst_compare};][]{2000ApJ...532..323P}.  
{So although we cannot rule out a contribution from the accreting intermediate mass black hole at the center of NGC 404, we conclude that a single SNa could be responsible for the shock-excited LINER emission at NIR wavelengths.} 
}
\end{itemize}

\acknowledgments
The authors would like to thank the anonymous referee for their helpful and constructive comments about this manuscript.  The data presented herein were obtained at the W. M. Keck Observatory, which is operated as a scientific partnership among the California Institute of Technology, the University of California and the National Aeronautics and Space Administration. The Observatory was made possible by the generous financial support of the W. M. Keck Foundation.  Finally, the authors wish to recognize and acknowledge the very significant cultural role and reverence that the summit of Maunakea has always had within the indigenous Hawaiian community.  We are very thankful to have had the opportunity to conduct observations from this mountain.

{\it Facilities:} \facility{W.M. Keck Observatory}.

\bibliographystyle{apj}
\bibliography{mybib_liners}{}

\clearpage

\end{document}